\documentclass{article}
\usepackage{graphicx} 
\usepackage{amsmath}
\usepackage{booktabs}
\usepackage{braket}
\usepackage{pifont}
\usepackage{epstopdf}
\usepackage{fancyhdr}
\usepackage{plain}
\usepackage{csquotes}
\usepackage{titletoc}
\usepackage[singlelinecheck=false]{caption}
\usepackage{amssymb}
\usepackage{amsmath}
\usepackage{multirow}
\usepackage{tablefootnote}
\usepackage[table]{xcolor}
\usepackage[export]{adjustbox}

\usepackage{caption}
\usepackage{subfigure}
\usepackage{amsfonts}
\usepackage{amssymb}
\usepackage{amsmath}
\usepackage{booktabs}
\usepackage{siunitx}
\usepackage{tabularx}
\usepackage{longtable}

\usepackage{float}
\usepackage{url}
\usepackage{cite}

\usepackage{todonotes}
\usepackage{import}
\usepackage{ragged2e}

\usepackage[subfigure]{tocloft}
\title{Feature Prediction in Quantum Graph Recurrent Neural Networks with Applications in Information Hiding}
\usepackage[a4paper, margin=1in]{geometry}

\author{
  Jawaher Kaldari\thanks{Jawaher Kaldari is with the Qatar Center for Quantum Computing, College of Science and Engineering, Hamad Bin Khalifa University, Doha, Qatar. E-mail: \texttt{jaka51804@hbku.edu.qa}} 
  \and 
  Saif Al-Kuwari\thanks{Saif Al-Kuwari is with the Qatar Center for Quantum Computing, College of Science and Engineering, Hamad Bin Khalifa University, Doha, Qatar. E-mail: \texttt{smalkuwari@hbku.edu.qa}}
}

\date{}
\begin{document}

\maketitle

\begin{abstract}
Graphs are a fundamental representation of complex, nonlinear structured data across various domains, including social networks and quantum systems. Quantum Graph Recurrent Neural Networks (QGRNNs) have been proposed to model quantum dynamics in graph-based quantum systems, but their applicability to classical data remains an open problem. 
In this paper, we leverage QGRNNs to process classical graph-structured data.
In particular, we demonstrate how QGRNN can reconstruct node features in classical datasets. Our results show that QGRNN achieves high feature reconstruction accuracy, leading to near-perfect classification. 
Furthermore, we propose an information hiding technique based on our QGRNN, where messages are embedded into a graph, then retrieved under certain conditions. We assess retrieval accuracy for different dictionary sizes and message lengths, showing that QGRNN maintains high retrieval accuracy, with minor degradation as complexity increases. 
These findings demonstrate the scalability and robustness of QGRNNs for both classical data processing and secure information hiding, paving the way for quantum-enhanced feature extraction, privacy-preserving computations, and quantum steganography.
\end{abstract}

\section{Introduction}

Breakthroughs in fields like image recognition and natural language processing have been driven by various neural network architectures, which have revolutionized the way we analyze and process both structured and unstructured data. While these models are excellent at extracting patterns from grid-like data structures, such as images, and from sequential data, such as text, they face serious challenges in tackling real-world problems that involve data represented as graphs. 
Graphs play a fundamental role in representing complex relationships in various domains, from social networks \cite{guo2025counterfactual} and transportation networks \cite{rahmani2023graph} to recommendation systems \cite{wu2022graph}. They model entities and their interactions, enabling a deeper understanding of complex systems. Graphs consist of nodes (also known as vertices) and edges, with the nodes representing entities and the edges representing connections or relationships between those entities. Graph Neural Networks (GNNs) have been widely recognized as powerful tools for analyzing graph-structured data, excelling in capturing relationships and dependencies within such structures \cite{wu2020comprehensive}. 
When graph structures incorporate time-varying data associated with their nodes or edges, they are commonly referred to as graph processes \cite{9054574}. For example, weather station networks can be modeled as graph processes, where hourly temperature readings are used to predict temperatures for the next hour \cite{perraudin2017stationary}. Graph Recurrent Neural Networks (GRNNs) are designed to model such time-varying data embedded within graph processes. 

Although classical computers excel in addressing a wide range of problems, they struggle when it comes to simulating systems governed by quantum mechanical properties. Systems exhibiting phenomena such as superposition and entanglement are difficult to model accurately using classical machine learning techniques. 
To address this challenge, quantum-inspired algorithms have been developed, leveraging quantum information principles while utilizing classical computing. While these algorithms provide improvements in specific scenarios, they do not fully harness the advantages of quantum mechanics. On the other hand, Quantum Machine Learning (QML) algorithms inherently utilize quantum principles to perform computations, enabling computations that supersede classical methods. 
Among the many areas where QML demonstrates significant potential is the analysis of graph-structured data, such as Quantum Graph Neural Networks (QGNNs) \cite{verdon2019quantum, beer2023quantum}, and Quantum Graph Kernel methods \cite{bai2017quantum}. 
One of the tasks in this domain is to learn the Hamiltonian of quantum systems represented as graphs. The Hamiltonian plays a crucial role in understanding the quantum dynamics of the system and the interactions between qubits. Quantum Graph Recurrent Neural Networks (QGRNNs) have been proposed as a powerful tool to learn the Hamiltonian of quantum systems that are represented as graphs \cite{verdon2019quantum}. What sets QGRNNs apart is their ability to predict the Hamiltonian without requiring direct access to the graph structure. Instead, QGRNNs leverage the time-evolved states of the system and its low-energy states to make predictions. Furthermore, QGRNNs efficiently utilize quantum resources, making them compatible with the current Noisy Intermediate-Scale Quantum (NISQ) devices \cite{preskill2018quantum}, demonstrating the advantages of these models without the need for fault-tolerant quantum computers. 


\subsection{Contribution}
In this paper, we shift the traditional use of QGRNN from understanding quantum dynamics to a novel approach for analyzing and processing classical graphs, where classical information is embedded directly into a graph structure. We then use classical machine learning to evaluate the performance of our model on two different datasets. We further demonstrate the utility of this approach by proposing an information hiding scheme based on our QGRNN, enabling secure embedding and retrieval of data.

\subsection{Organization}
The remainder of this paper is organized as follows. Section \ref{sec:Preliminaries} provides background on relevant concepts necessary for understanding QGRNN,  including a detailed description of QGRNNs. In Section \ref{sec:methodology}, we introduce our methodology, followed by the results presented in Section \ref{sec:results}. Section \ref{sec:applications} explores a novel application of our methodology, and finally, Section \ref{sec:conclusion} provides the conclusion.

\section{Preliminaries}\label{sec:Preliminaries}
In this section, we provide a brief background on both the classical and quantum Ising models, as well as QGRNN.

\subsection{Classical Ising Model}
The (classical) Ising model provides a framework for understanding magnetic materials by representing them as systems of molecules arranged on a lattice graph with nearest-neighbor interactions \cite{bian2010ising, brush1967history}. Each molecule has a spin that can be $\sigma_{i,j}= \pm 1$ relative to an applied magnetic field, and these spins interact with each other. Here, the material is represented by a two-dimensional square lattice, where $i$ and $j$ are the columns and rows of the square, respectively. Since it is a two-dimensional square lattice, each molecule is connected to four neighboring nodes,  and interactions occur only between adjacent nodes. The Ising model Hamiltonian is defined as follows \cite{cipra1987introduction, choi2021tutorial}:

\begin{equation}\label{ising}
H = -J \sum_{( i, j) } \sigma_i \sigma_j - B \sum_i h_i \sigma_i 
\end{equation}

\noindent where $J$ represents the interactions between neighboring molcules, $B$ is the magnetic momentum, and $h$ is the externally applied magnetic field.

\subsection{Transverse-field Ising Model}
The transverse-field Ising model (TIM) is a quantum extension of the classical Ising model. This model is particularly useful because it captures effects, such as quantum phase transitions, that classical Ising models fail to explain \cite{dutta2010quantum}. Unlike the classical model, where spins can only take discrete values of $-1$ or $1$, the TIM incorporates quantum phenomena, allowing a spin or a qubit to exist in a superposition state. The TIM Hamiltonian is defined as follows \cite{choi2021tutorial}:

\begin{equation}\label{TIMeq}
 H = J \sum_{(i, j )} \sigma_i^z \sigma_j^z - B \sum_i h_i \sigma_i^z - B\sum_i g_i \sigma_i^x
\end{equation}

\noindent where $\sigma_i^z$ and $\sigma_j^z$ are Pauli-Z operators acting on $i$ and $j$, respectively, while $\sigma_i^x$ is a Pauli-X operator acting on $i$. These Pauli operators are defined as follows:

\[
\sigma^z = 
\begin{pmatrix}
1 & 0 \\
0 & -1
\end{pmatrix}, \quad
\sigma^x = 
\begin{pmatrix}
0 & 1 \\
1 & 0
\end{pmatrix}.
\]

The parameters $J$, $B$ and $h_i$ are as defined in equation \ref{ising}, and $g_i$ is the longitudinal magnetic field.

TIM enables us to calculate the Hamiltonian of a quantum system represented as a graph or a lattice, where the nodes correspond to qubits and the edges represent the interactions between these qubits.

\subsection{Quantum Graph Recurrent Neural Network}\label{sec:QGRNN}

Quantum Graph Recurrent Neural Network (QGRNN) was proposed in \cite{verdon2019quantum} and is designed to model real-time Hamiltonian dynamics in graph-based quantum systems. Like classical recurrent neural networks, QGRNNs share parameters through the sequential application of recurrent layers. QGRNNs are very effective in scenarios where the graph's structure, including its nodes and edges, is unknown or inaccessible. It uses the low-energy states and time-evolved states of the system to predict the Hamiltonian parameters.


The definition of QGRNN is derived from the TIM equation (equation \ref{TIMeq}), which can be further simplified to \cite{choi2021tutorial}:

\begin{equation}\label{QGRNNeq}
    H (\alpha)= {\alpha_{i,j}}^{(1)} \sum_{(i, j )} \sigma_i^z \sigma_j^z + {\alpha_{i}}^{(2)}  \sum_i  \sigma_i^z +\sum_i  \sigma_i^x 
\end{equation}

\noindent where $\alpha^{(1)}$ corresponds to the weights of the edges of the graph, which represent the interactions between the qubits, while $\alpha^{(2)}$ corresponds to the weights of the nodes or qubits. 
Equation \ref{QGRNNeq} serves as the basis for constructing the ansatz of the QGRNNs. Using this Hamiltonian, the time-evolution unitary operator can be expressed as:

\begin{equation}
U = e^{-itH} 
\end{equation}

\noindent Substituting $H(\alpha)$ into the unitary operator, we obtain:

\begin{equation}\label{eq:unitary}
    U = e^{-itH(\alpha)} = e^{-it\left({\alpha_{i,j}}^{(1)} \sum_{(i, j )} \sigma_i^z \sigma_j^z + {\alpha_{i}}^{(2)}  \sum_i  \sigma_i^z +\sum_i  \sigma_i^x \right)} =e^{-it(H_1(\alpha)+H_2(\alpha))}
\end{equation}

\noindent where $H_1(\alpha)={\alpha_{i,j}}^{(1)} \sum_{(i, j )} \sigma_i^z \sigma_j^z + {\alpha_{i}}^{(2)}  \sum_i  \sigma_i^z $ and $H_2(\alpha)=\sum_i  \sigma_i^x$. By separating $H(\alpha)$ into $H_1(\alpha)$ and $H_2(\alpha)$, we can approximate the time-evolution operator using methods like Trotter decomposition \cite{broughton2020tensorflow}. This step is crucial because the terms in Equation \ref{eq:unitary} are non-commuting, making them challenging to execute directly on quantum computers \cite{jones2019optimising}. The Trotter-Suzuki formula is given by \cite{Ceroni_2020}:

\begin{equation}\label{eq:trotter}
  e^{a + b } \approx \left(e^{a/n} e^{b/n} \right)^n
\end{equation}

\noindent Substituting the Hamiltonian from Equation \ref{eq:unitary} into the Trotter-Suzuki formula, we approximate the time-evolution operator as:

\begin{equation}
  e^{-it(H_1 + H_2)} \approx \left(e^{-i\delta H_1} e^{-i\delta H_2} \right)^{D}
\end{equation}

\noindent where $D$ represents the total number of Trotter steps and is given by $D=\frac{t}{\delta}$, with $\delta$ being the Trotter step size.
The definition of the QGRNN anzats is based on the time-evolution operator, expressed as:
\begin{equation}\label{ansatz}
   U_H(\delta,\alpha)= \prod_{k=1}^{D} \left( \prod_{j=1}^{Q} e^{-i\delta_{j} {H}_{j}(\alpha)}\right)
\end{equation}

QGRNN is expressed as a repetitive sequence of exponentials of graph Hamiltonians, where $Q$ denotes the sequence of different Hamiltonian evolutions, and the whole sequence is repeated $D$ times. This particular QGRNN ansatz is used to learn the Hamiltonian of quantum systems. 

\begin{figure}
    \centering
    \includegraphics[width=0.65\linewidth]{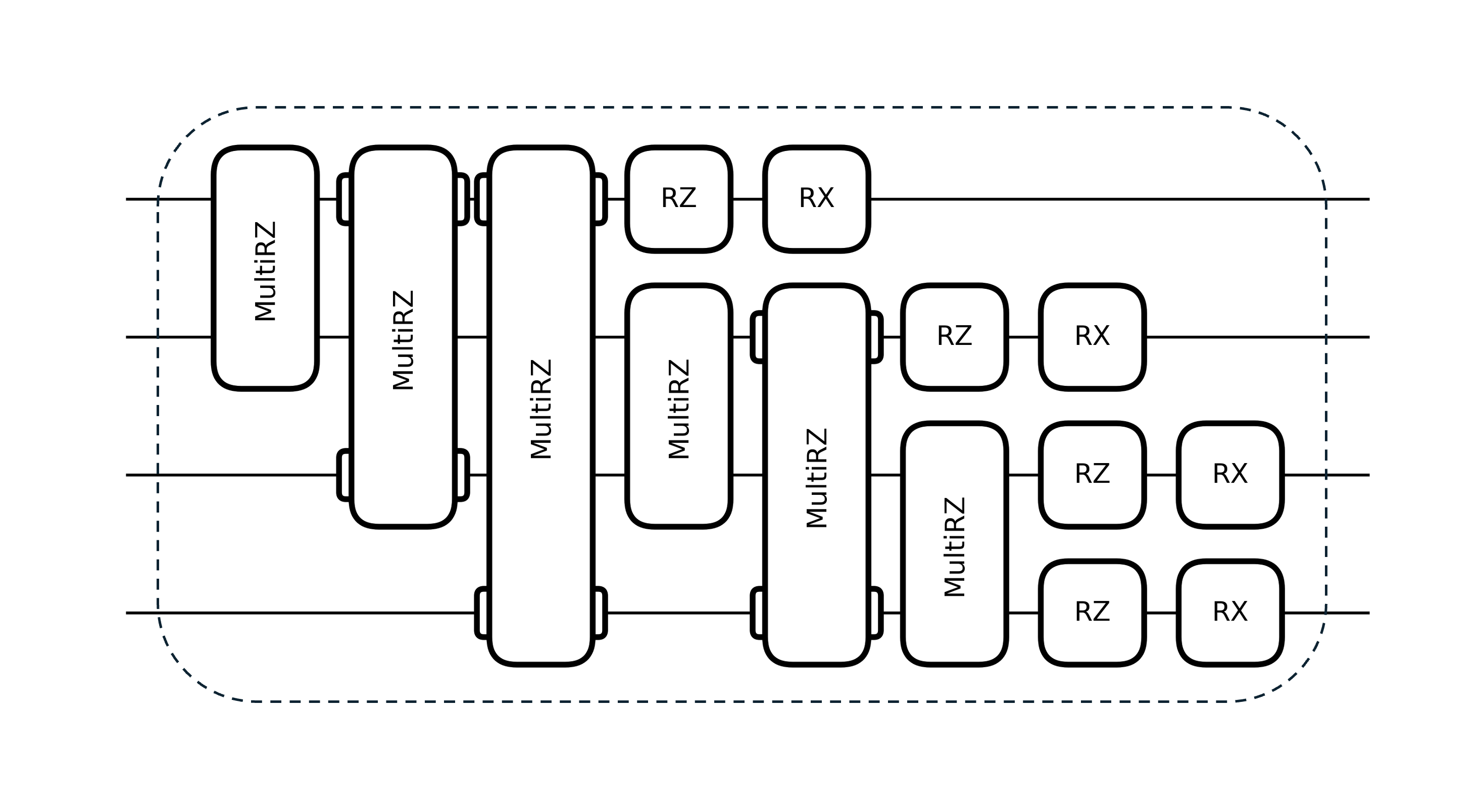}
    \caption{QGRNN layers for processing a 4-node graph}
    \label{fig:qgrnn_layers}
\end{figure}

Figure \ref{fig:qgrnn_layers} illustrates the QGRNN layers (for a 4-node graph) that are derived from the ansatz in Equation \ref{ansatz}. These layers are repeated according to $D=\frac{t}{\delta}$ and implemented using a combination of multi-qubit Pauli-Z rotation gates, single-qubit Pauli-Z and Pauli-X rotation gates, as described in Equation \ref{ansatz}. 

\subsubsection{Training}\label{sec:training_process}
Consider a scenario in which a quantum system is governed by an unknown Hamiltonian $H_{target}(\theta)$, where $\theta=\{ \theta^1,\theta^2\}$ represents the parameters of the system. However, neither the parameter $\theta$ nor the graph structure of the system is known. The only available information about the system consists of the low-energy states $\ket{\psi_0}$ of the system and the time-evolved states $\ket{\psi_t}=\{ \ket{ \psi(t_1)}, \ket{\psi(t_2)}, \ldots, \ket{\psi(t_N)} \}$, which are defined as:

\begin{equation}\label{time-evolvedstates}
\ket{\psi_t}\equiv U(t) \ket{\psi_0}= e^{-i t H_{target}(\theta)}
\end{equation}

\noindent where $t \in [0, t_{max}]$ is random but known. Our goal is to use QGRNN to learn the parameters of the Hamiltonian equation: 

\begin{equation}
    H_{target}(\theta)= {\theta_{i,j}}^{(1)} \sum_{(i, j )} \sigma_i^z \sigma_j^z + {\theta_{i}}^{(2)}  \sum_i  \sigma_i^z +\sum_i  \sigma_i^x 
\end{equation}

To learn the Hamiltonian, we start with a graph initialized with random parameters $\beta=\{ \beta^1,\beta^2\}$. We then apply the QGRNN layers to the low-energy states we have $U_H(\delta,\beta)\ket{\psi_0}$. Then the parameters $\beta=\{ \beta^1,\beta^2\}$ are optimized to increase the similarity between the time-evolved states and the states prepared using the QGRNN:
\[
U_H(\delta,\beta) \ket{\psi_0} \approx \ket{\psi_t} 
\]

As the states converge, $\beta \rightarrow \theta$, indicating that the target Hamiltonian parameters have been successfully learned. 

\subsubsection{Cost Function}
\begin{figure}
    \centering
    \includegraphics[width=0.4\linewidth]{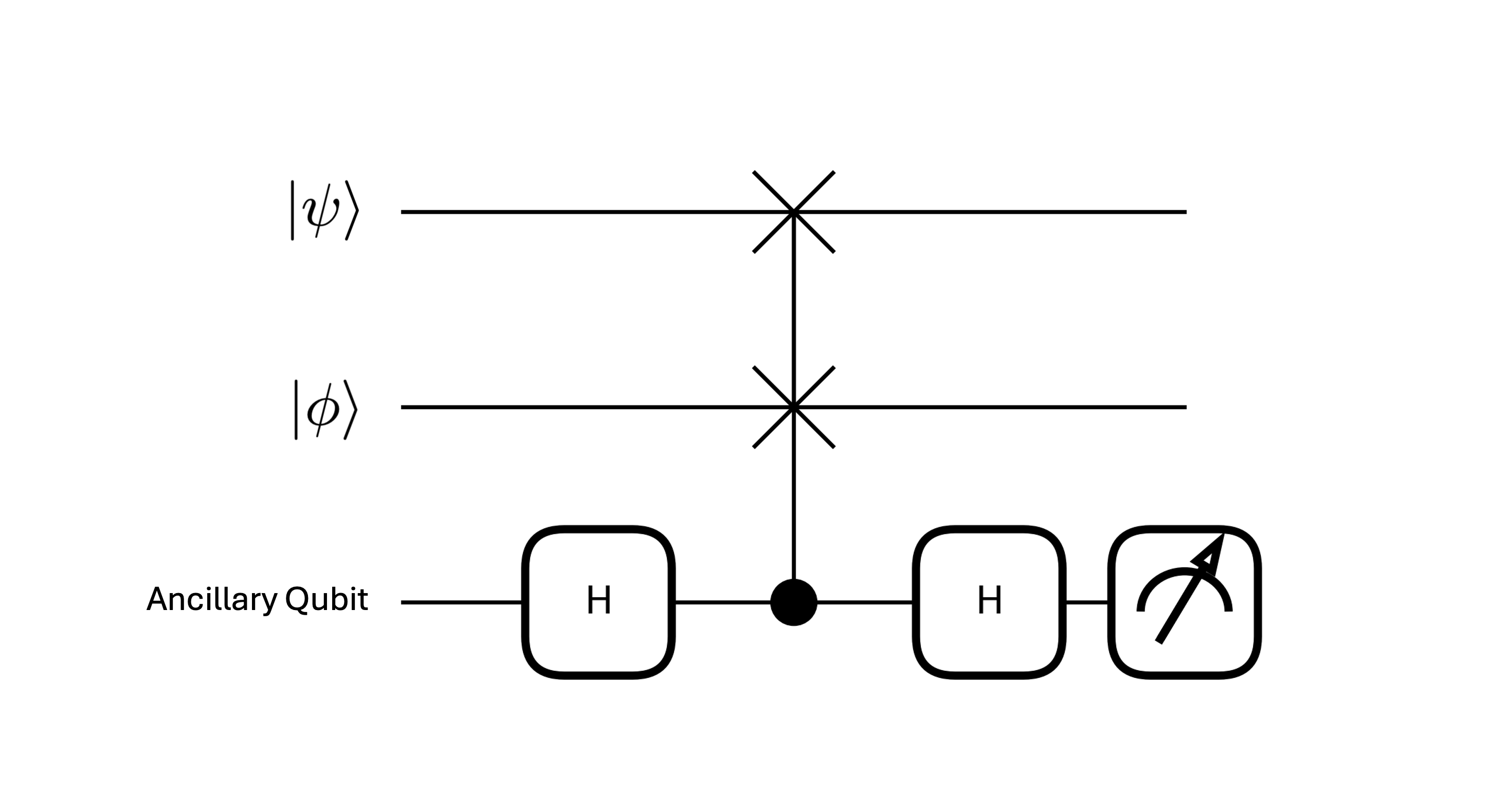}
    \caption{SWAP test circuit}
    \label{fig:SWAP_test}
\end{figure}

Learning the Hamiltonian parameters is based on maximizing the similarity between the two states, $U_H(\delta,\beta)\ket{\psi_0}$ and 
$\ket{\psi_t}$. To do this, we need to be able to quantify their similarity without directly measuring the qubits, which can be achieved using the SWAP test \cite{fanizza2020beyond}. This forms the basis for the cost function in QGRNN. The SWAP test is used to compute the fidelity between two states, which quantifies how close they are to each other. The test involves using a controlled-SWAP gate, two Hadamard gates, and an ancillary qubit. Given two quantum states $\ket{\psi}$ and $\ket{\phi}$, the goal is to measure their fidelity, which is given by $F = |\langle \psi|\phi\rangle|^2$ \cite{ripper2022swap}. The SWAP test circuit used to measure the similarity is shown in Figure \ref{fig:SWAP_test}. 

For QGRNN, the goal is to measure and maximize the similarity between the states $U_H(\delta,\beta)\ket{\psi_0}$ and $\ket{\psi_t}$. Using the SWAP test, the fidelity becomes $F=\left| \langle \psi_t \, | \, U_H(\delta, \beta) \, | \, \psi_0 \rangle \right|^2$. As the states become more similar, this expression approaches $1$, while approaching $0$ as they diverge.
To formulate this problem as a cost function, the problem becomes minimizing $ - \left| \langle \psi_t \, | \, U_H(\delta, \beta) \, | \, \psi_0 \rangle \right|^2$. Since we have a batch of quantum data, the cost function becomes the average negative fidelity, expressed as:

\begin{equation}
 L(\beta, \delta) = -\frac{1}{N} \sum_{i=1}^N \left| \langle \psi(t_i) | U_H(\beta, \delta) | \psi_0 \rangle \right|^2   
\end{equation}

\noindent where $N$ is the number of quantum states processed, or the batch size.

\section{Methodology}\label{sec:methodology}
Our proposal leverages QGRNN to predict and reconstruct hidden features of a dataset represented as a classical graph, relying solely on time-evolved states when direct access to the graph is unavailable. The reconstructed features are then used as input for classical machine learning models to perform classification tasks. To validate this approach, we apply QGRNN to two well-known classical datasets: Iris  \cite{fisher1936use}, and MNIST \cite{lecun2002gradient}. Features from each dataset are embedded into the nodes of a graph. Since direct access to the graph is unavailable, we simulate its time evolution using the Hamiltonian of the graph, generating time-evolved states that encode the underlying structure. These states serve as the only available information related to the graph. 
QGRNN is then trained to reconstruct the hidden node features and the original graph structure from the time-evolved states. The reconstructed features are subsequently used as input for classical machine learning models to evaluate the effectiveness of the QGRNN-based reconstruction.


\subsection{Preprocessing and Dimensionality Reduction}
The Iris dataset consists of four numerical features and does not require preprocessing, as its compact size makes it directly compatible with QGRNN. However, the feature values were scaled to the range $[0, 5]$ to improve training stability, as QGRNN learns better when feature values are within a controlled range. 

Unlike Iris, MNIST dataset does require preprocessing. MNIST images consist of 28x28 pixels, which are challenging to represent in a graph structure for processing by QGRNN due to the limited number of qubits available in current NISQ devices. 
Since the number of nodes in the graph corresponds to the number of qubits, directly encoding all 784 pixels as nodes would require a quantum system with a prohibitively large number of qubits, which is currently infeasible. Furthermore, the total number of qubits required is not equal to the number of nodes, but follows a specific structure that includes additional registers for state preparation (see Section \ref{sec:xperimental Setup and training parameters}). 
To address this problem, we apply Principal Component Analysis (PCA) to reduce the dimensionality of the MNIST dataset. PCA transforms the original 784-dimensional pixel space into a lower-dimensional space while preserving the most relevant information. After PCA is performed, we retain six principal components that represent the most important features of the images.  To ensure consistency with QGRNN’s optimization process, these PCA components were also scaled in the range [0, 5], similar to the Iris dataset.

\subsection{Graph Representation}
In this setup, we embedded data into the nodes of a graph, with each node representing a specific component or feature. For the MNIST dataset, we embed the six principal components obtained from PCA into six nodes. For the Iris dataset, the features are embedded directly into the nodes. 
Edges were not explicitly used in this setup and were instead initialized randomly -- for this proof-of-concept demonstration we choose to focus on the nodes for simplicity. However, incorporating edge information could provide additional space for data encoding and redundancy in future applications.

\subsection{Time Evolution Simulation }

To simulate the time-evolved states, we follow these steps:
\begin{enumerate}
    \item We begin with randomly initialized quantum states $\ket{\psi_{initial}}$, which serve as starting states for the evolution \footnote{In QGRNN, low-energy states are typically used. These states are usually simulated using Variational Quantum Eigensolver (VQE),  as only such states are assumed to be accessible in practical quantum systems. However, since we simulate classical graphs and have full knowledge of the system, we simplify the process by using randomly initialized quantum states, which we precompute and reuse. Despite this simplification, QGRNN remains effective.}.
    \item The Hamiltonian of the graph is computed using the Transverse Field Ising Model, as described in Equation \ref{QGRNNeq}
    \item Using this Hamiltonian, the system undergoes time evolution simulation, generating a set of time-evolved states.
\end{enumerate}

Both the randomly initialized states and the time-evolved states are used as input to QGRNN, which then predicts the nodes of the graph, corresponding to the PCA components of the MNIST images or the features of the Iris dataset, given that we do not have access to the original graph.

\subsection{Classification}
Once QGRNN predicts the nodes of the graph, the next step is to classify the features reconstructed from these predictions. To evaluate the effectiveness of QGRNN-based feature extraction, we use multiple classical machine learning models. The classifiers are trained to perform classification based on the six PCA components for the MNIST dataset and the original features for the Iris dataset. 
We compare the performance of different classifiers, including:
Logistic Regression, Decision Tree, Random Forest, Naïve Bayes and Support Vector Machine.




\section{Results}\label{sec:results}
Our goal is to evaluate the ability of QGRNN to reconstruct hidden node features of classical datasets using only time-evolved states and the randomly initialized states. 

\subsection{Experimental Setup}\label{sec:xperimental Setup and training parameters}
In our setup, two quantum registers and an ancillary qubit are required. Given a graph with $n$ nodes, the total number of qubits needed is $2n+1$, where, the first register ($n$ qubits) is used to prepare the time-evolved states, the second register ($n$ qubits) is used to apply QGRNN layers to the randomly initialized states, which are compared to the time-evolved states, and the ancillary qubit (1 qubit) is used for the SWAP test, which is measured to evaluate the similarity between states.

The training process of QGRNN, including the optimization of Hamiltonian parameters and the application of the cost function, follows the methodology outlined in Section \ref{sec:training_process}. The ansatz structure is implemented as described in that section. The performance of QGRNN depends on key hyperparameters, including the number of Trotter steps, learning rate, optimization method, and other relevant factors. Table \ref{tab:hyperparams} summarizes the hyperparameters used. It is worth mentioning that although the learning rate of $0.5$ is relatively high, it led to stable convergence in fewer training epochs, making it an efficient and effective choice in our setup.

 \begin{table}[h!]
\centering
\caption{Hyperparameters Used in QGRNN Training}
\begin{tabular}{ll}
\toprule
\textbf{Hyperparameter} & \textbf{Value} \\ 
\midrule
Batch Size & 15 \\ 
Learning Rate & 0.5 \\ 
Number of Epochs & 150 \\ 
Trotter Step Size &  0.01\\ 
Maximum Evolution Time ($t_{\text{max}}$) & 0.5 \\
Optimizer & Adam Optimizer \\ \bottomrule
\end{tabular}
\label{tab:hyperparams}
\end{table}

\subsection{Evaluation Metrics}
To evaluate the accuracy of QGRNN's predictions and performance, we employ the following evaluation metrics:

\begin{itemize}
\item \textbf{Mean Squared Error (MSE):} MSE measures the average squared difference between predicted values and actual (true) values. It is defined as:

\[
\text{MSE} = \frac{1}{N} \sum_{i=1}^{N} (y_i - \hat{y}_i)^2
\]

wher $y_i$ is the actual value, $\hat{y}_i$ is the predicted value, and $N$ is the number of samples. 

\item \textbf{Root Mean Squared Error (RMSE):} RMSE is the square root of the MSE, and is defined as: 

\[
\text{RMSE} = \sqrt{ \frac{1}{N} \sum_{i=1}^{N} (y_i - \hat{y}_i)^2 }
\]

\item \textbf{Mean Absolute Error (MAE):} MAE measures the average magnitude of the errors. The formula to calculate MAE is:

\[
\text{MAE} = \frac{1}{N} \sum_{i=1}^{N} \left| y_i - \hat{y}_i \right|
\]

\item \textbf{Cosine Similarity:} Cosine similarity measures the similarity between two non-zero vectors by calculating the cosine of the angle between them. This reflects how closely the vectors align in direction. It can be calculated using: 

\[
\text{Cosine Similarity} = \frac{\vec{y} \cdot \vec{\hat{y}}}{\|\vec{y}\| \, \|\vec{\hat{y}}\|}
\]

\end{itemize}
We evaluate the performance of our QGRNN using two datasets: MNIST (PCA-reduced) and Iris. The evaluation consists of two steps. 

\begin{enumerate}
    \item Feature Prediction: The accuracy of QGRNN’s predicted node features is measured using the metrics detailed above.
    \item Classification evaluation: The predicted features are used as input for classical classification models, and their performance is evaluated. 
\end{enumerate}

\subsection{Feature Prediction}
In this section, we compare the QGRNN predicted node values with the actual dataset features. To assess the accuracy of QGRNN’s predictions, we use the following evaluation metrics: MSE, RMSE, MAE and Cosine Similarity. To further analyze QGRNN’s training behavior, we present the cost function plots. Additionally, we compare the Hamiltonians derived from the predicted nodes and edges with the corresponding target Hamiltonians as a validation metric. We first evaluate QGRNN's performance on the Iris dataset, followed by MNIST.

\subsubsection{Results on Iris}
The Iris dataset consists of four features per sample point. The objective is to reconstruct these features that were originally embedded into the graph from the time-evolved states and $\ket{\psi_{initial}}$. The comparison of the actual and predicted features is presented in Table \ref{tab:iris_predictions}, which shows that QGRNN closely reconstructs the node features across all six samples.

\begin{table*}[h!]
\centering
\caption{Comparison of Actual vs. Predicted Features for the Iris Dataset}
\begin{tabular}{lll}
\toprule
\textbf{Sample} & \textbf{Actual Features} & \textbf{Predicted Features} \\ 
\midrule
 1 & (1.944, 3.75, 0.593, 0.417) & (1.961, 3.811, 0.601, 0.398) \\ 
2 & (1.528, 2.917, 0.424, 0.625) & (1.539, 2.972, 0.422, 0.610) \\ 
 3 & (2.500, 1.667, 3.136, 2.292) & (2.531, 1.654, 3.153, 2.315) \\ 
 4 & (2.083, 1.458, 2.458, 2.292) & (2.110, 1.450, 2.459, 2.316) \\ 
 5 & (4.722, 1.250, 5.000, 4.583) & (4.833, 1.296, 4.925, 4.514) \\ 
 6 & (3.611, 2.292, 3.475, 4.583) & (3.633, 2.309, 3.542, 4.644) \\ \bottomrule
\end{tabular}
\label{tab:iris_predictions}
\end{table*}

Table \ref{tab:iris_accuracy} further quantifies QGRNN's prediction performance.  The results indicate high prediction accuracy across all samples, as evidenced by low MSE, RMSE, and MAE values, as well as near-perfect Cosine Similarity scores. The MSE values range from 0.000345 to 0.006218, which indicates that the squared deviations between the actual and the predicted features are minimal. RMSE values, which are the square roots of MSE, range from 0.0186 to 0.0789, which is also considered low. The largest MSE and RMSE values (0.006218 and 0.0789, respectively) suggest that QGRNN has slightly more difficulty reconstructing certain features. However, these errors remain small, indicating that the model consistently performs well across all samples. The MAE values range from 0.0152 to 0.0754, suggesting that it is consistently low. Cosine Similarity values remain above 0.999 for all samples, indicating that QGRNN predictions closely match the actual features.

\begin{table}[h!]
\centering
\caption{Prediction Metrics for Each Sample in the Iris Dataset}
\begin{tabular}{lllll}
\toprule
\textbf{Sample} & \textbf{MSE} & \textbf{RMSE} & \textbf{MAE} & \textbf{Cosine Similarity} \\ \hline
1 & 0.001107 & 0.033274 & 0.026025 & 0.999979 \\ 
 2 & 0.000844 & 0.029043 & 0.020698 & 0.999961 \\
3 & 0.000491 & 0.022155 & 0.020992 & 0.999983 \\ 
4 & 0.000345 & 0.018586 & 0.015220 & 0.999979 \\ 
 5 & 0.006218 & 0.078855 & 0.075352 & 0.999823 \\ 
 6 & 0.002285 & 0.047803 & 0.042216 & 0.999988 \\ 
\bottomrule
\end{tabular}
\label{tab:iris_accuracy}
\end{table}

\begin{figure}[H]
    \centering
    \subfigure[Cost results for Sample 1]{\includegraphics[width=0.3\textwidth]{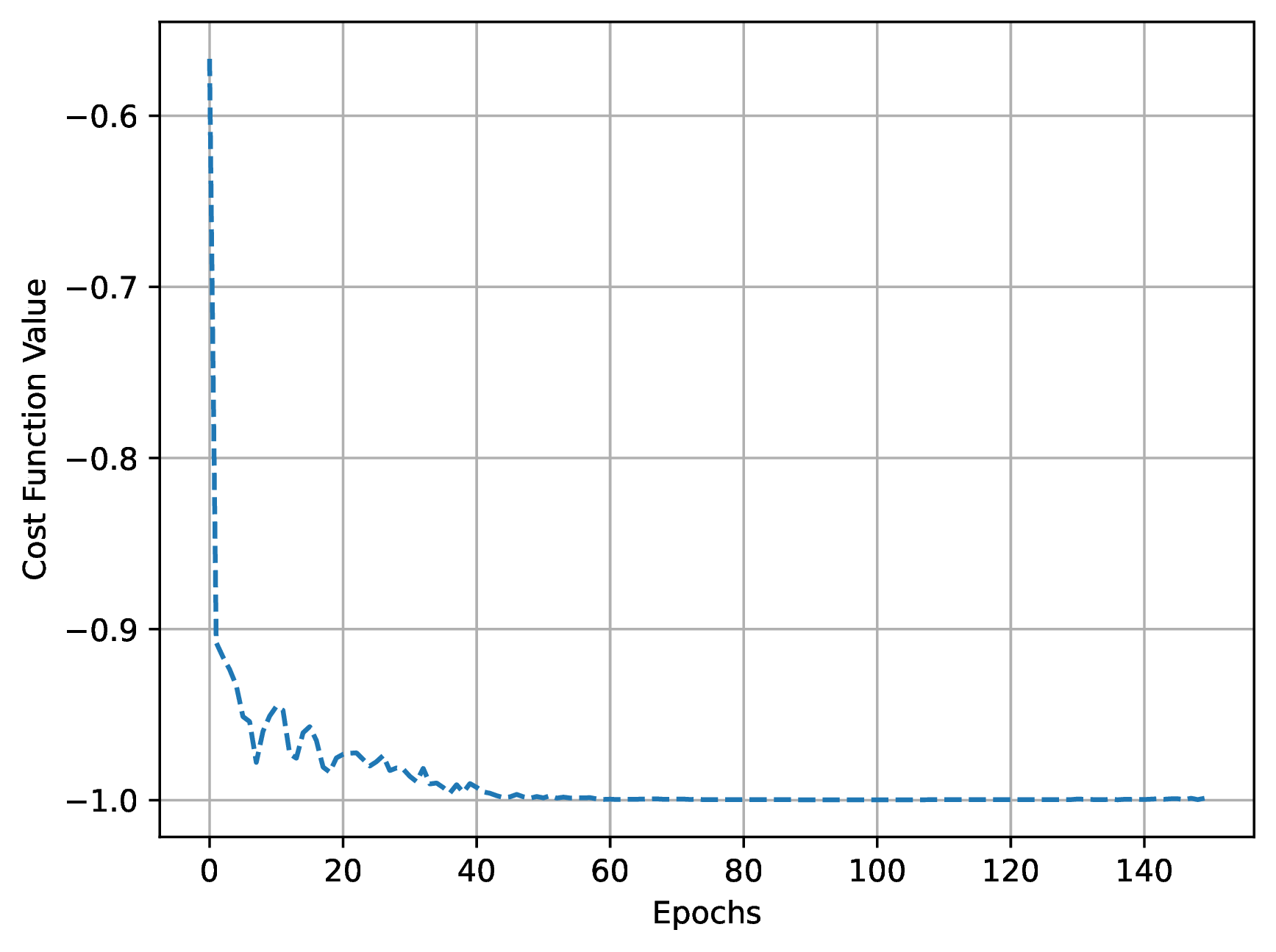}}
    \subfigure[Cost results for Sample 2]{\includegraphics[width=0.3\textwidth]{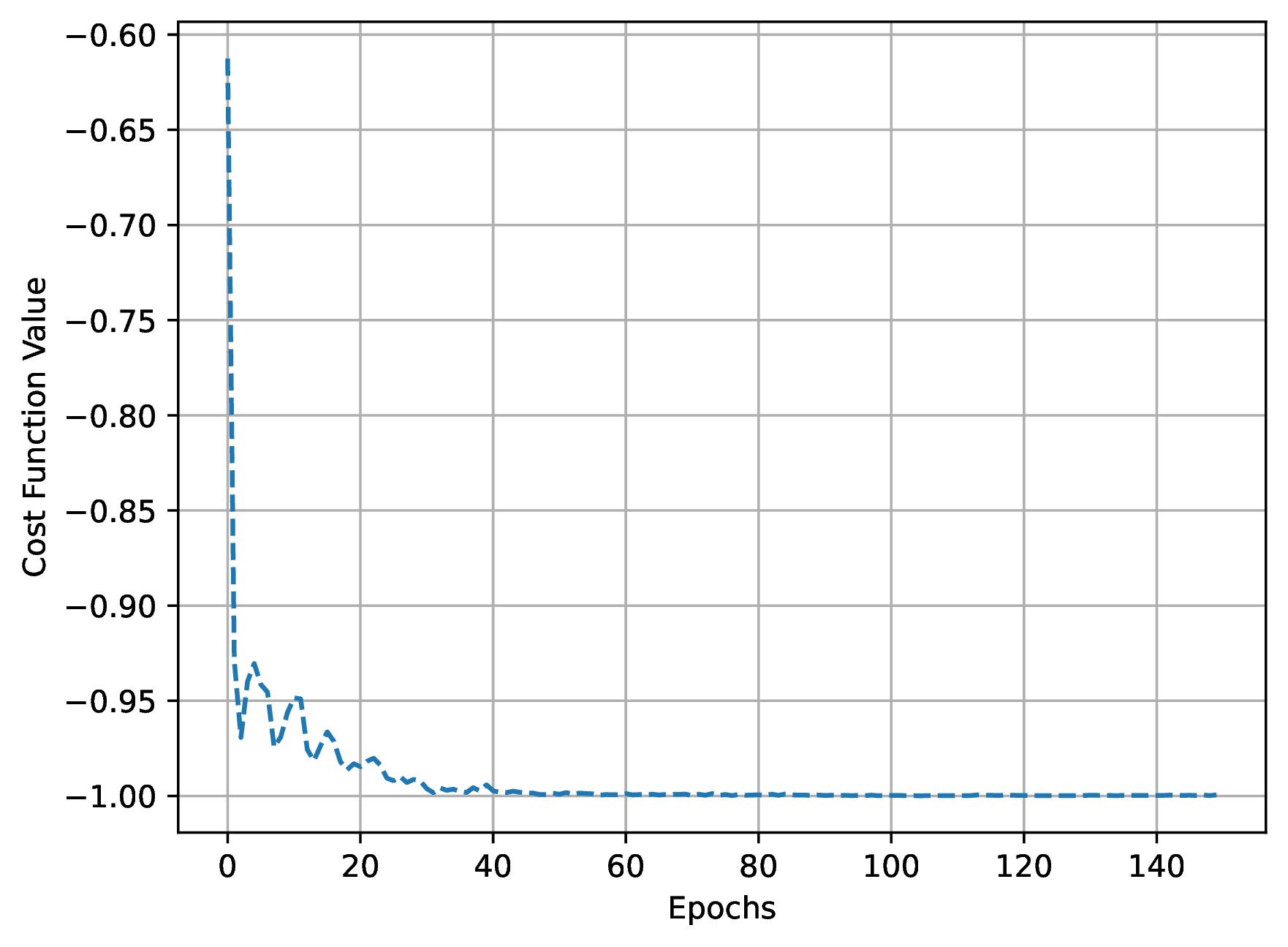}}
    \subfigure[Cost results for Sample 3]{\includegraphics[width=0.3\textwidth]{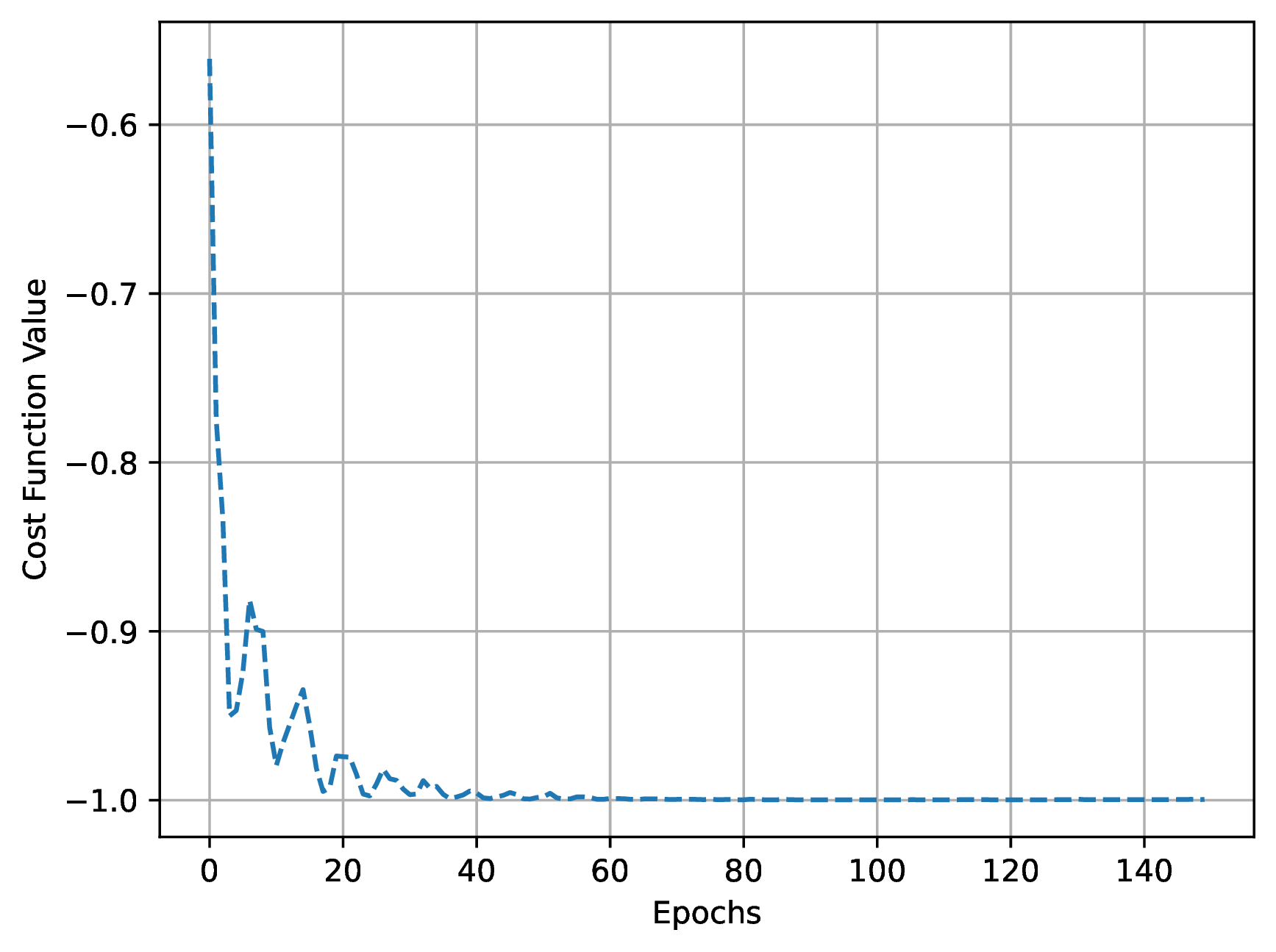}}
    \subfigure[Cost results for Sample 4]{\includegraphics[width=0.3\textwidth]{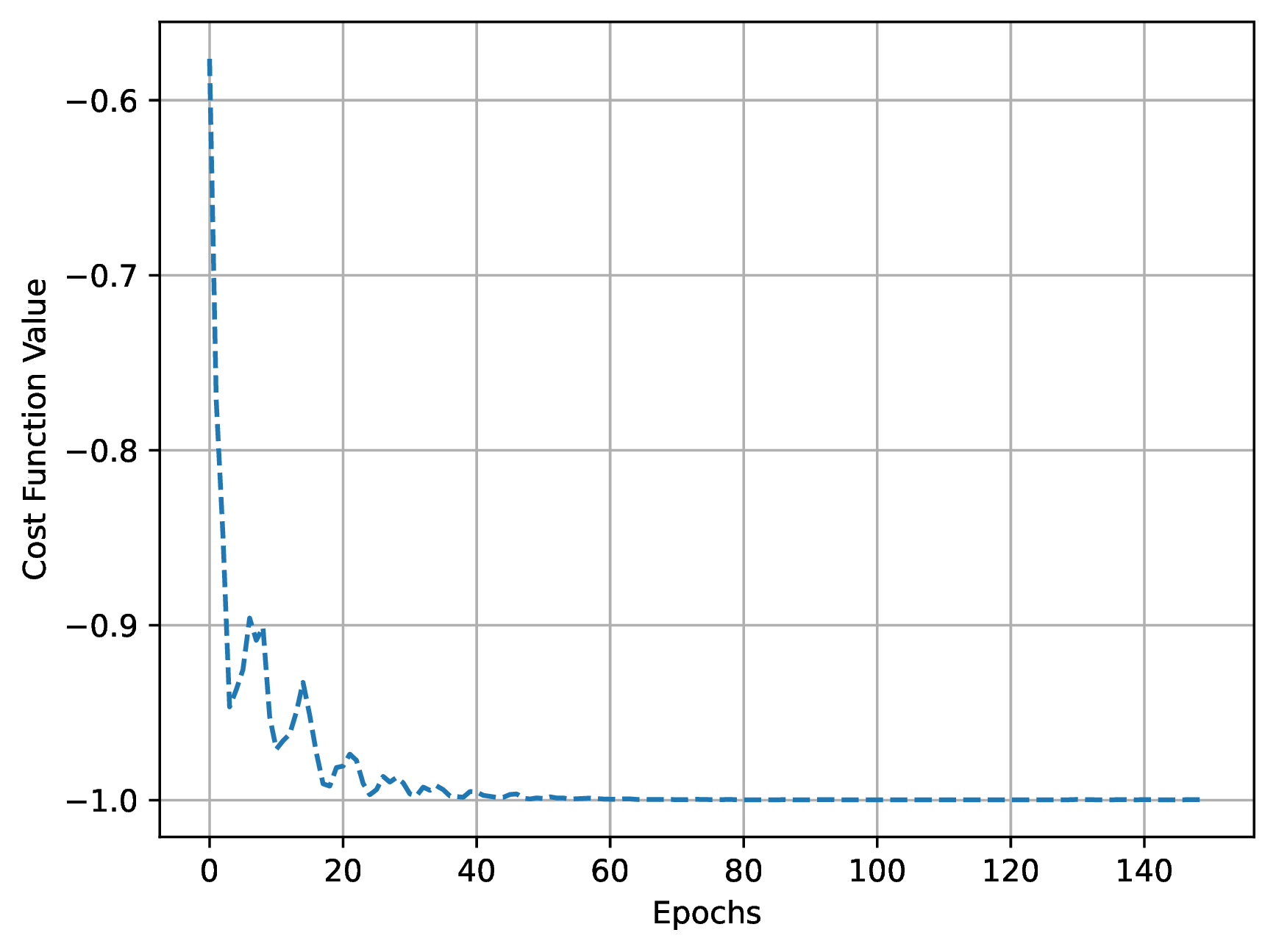}}
    \subfigure[Cost results for Sample 5]{\includegraphics[width=0.3\textwidth]{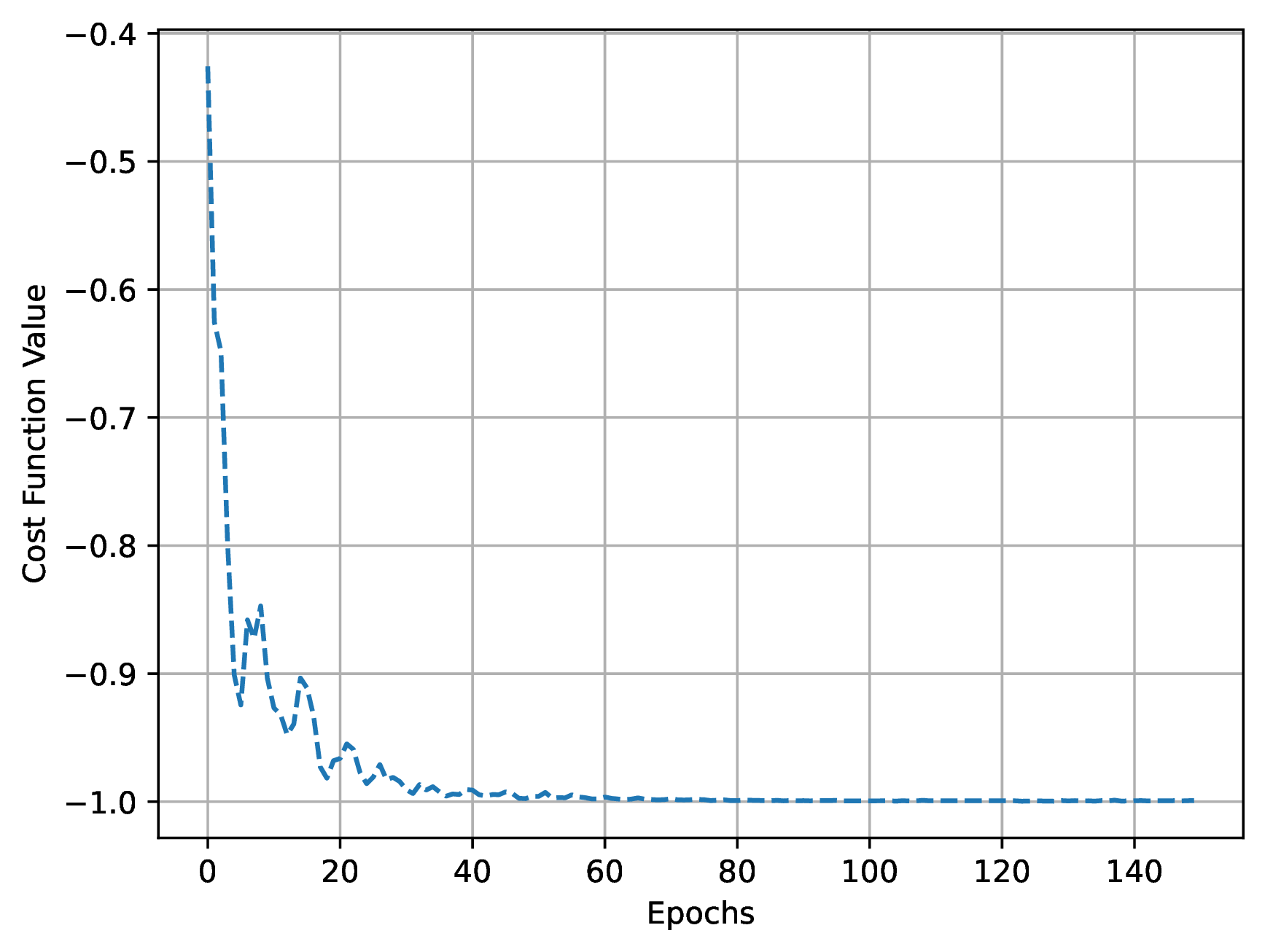}}
    \subfigure[Cost results for Sample 6]{\includegraphics[width=0.3\textwidth]{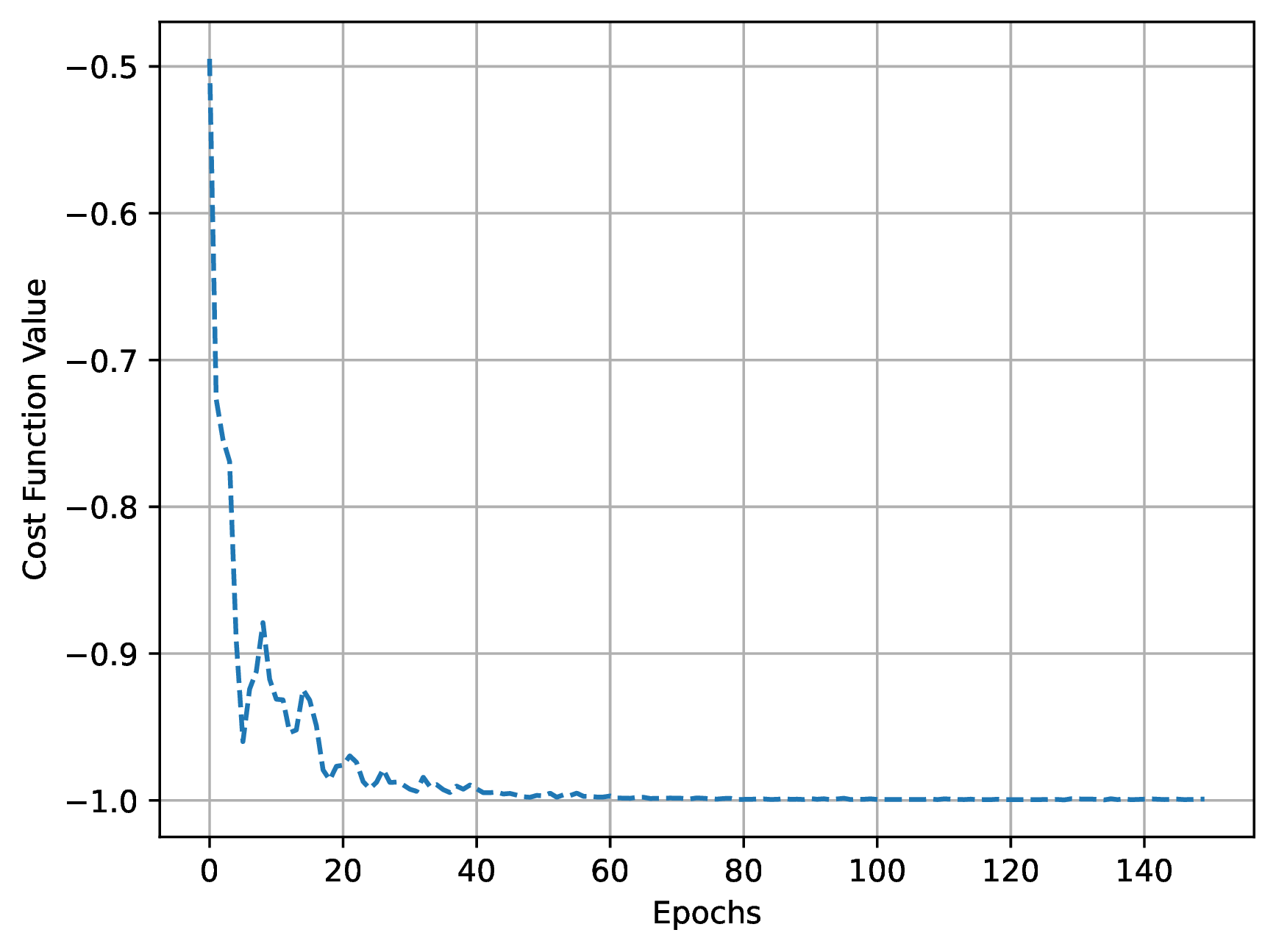}}  
    \caption{Cost function value for all tested Iris samples}
    \label{fig:cost_function_iris}
\end{figure}

During training, QGRNN optimizes the predicted node features to match the actual features by minimizing the cost function. Figure~\ref{fig:cost_function_iris} illustrates the cost function behavior for all six Iris samples over training iterations. The rapid decrease in the cost function demonstrates that QGRNN quickly learns the features. All samples exhibit a similar overall trend, demonstrating that QGRNN reliably learns the features from the time-evolved states and $\ket{\psi_{initial}}$.

\begin{figure}[H]
    \centering
    \subfigure[Target and learned Hamiltonian for Sample 1]{\includegraphics[width=0.49\textwidth]{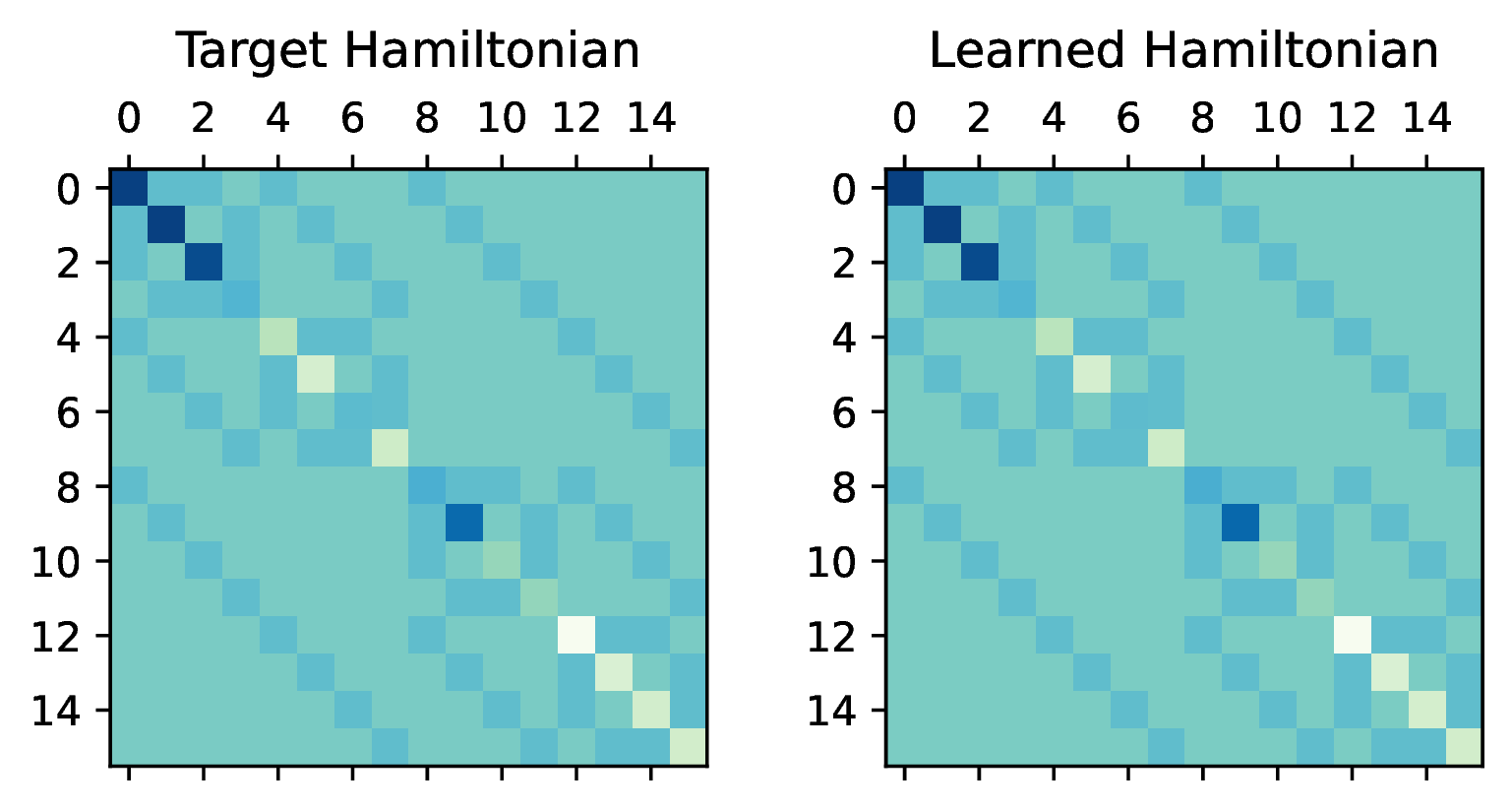}}
    \subfigure[Target and learned Hamiltonian for Sample 2]{\includegraphics[width=0.49\textwidth]{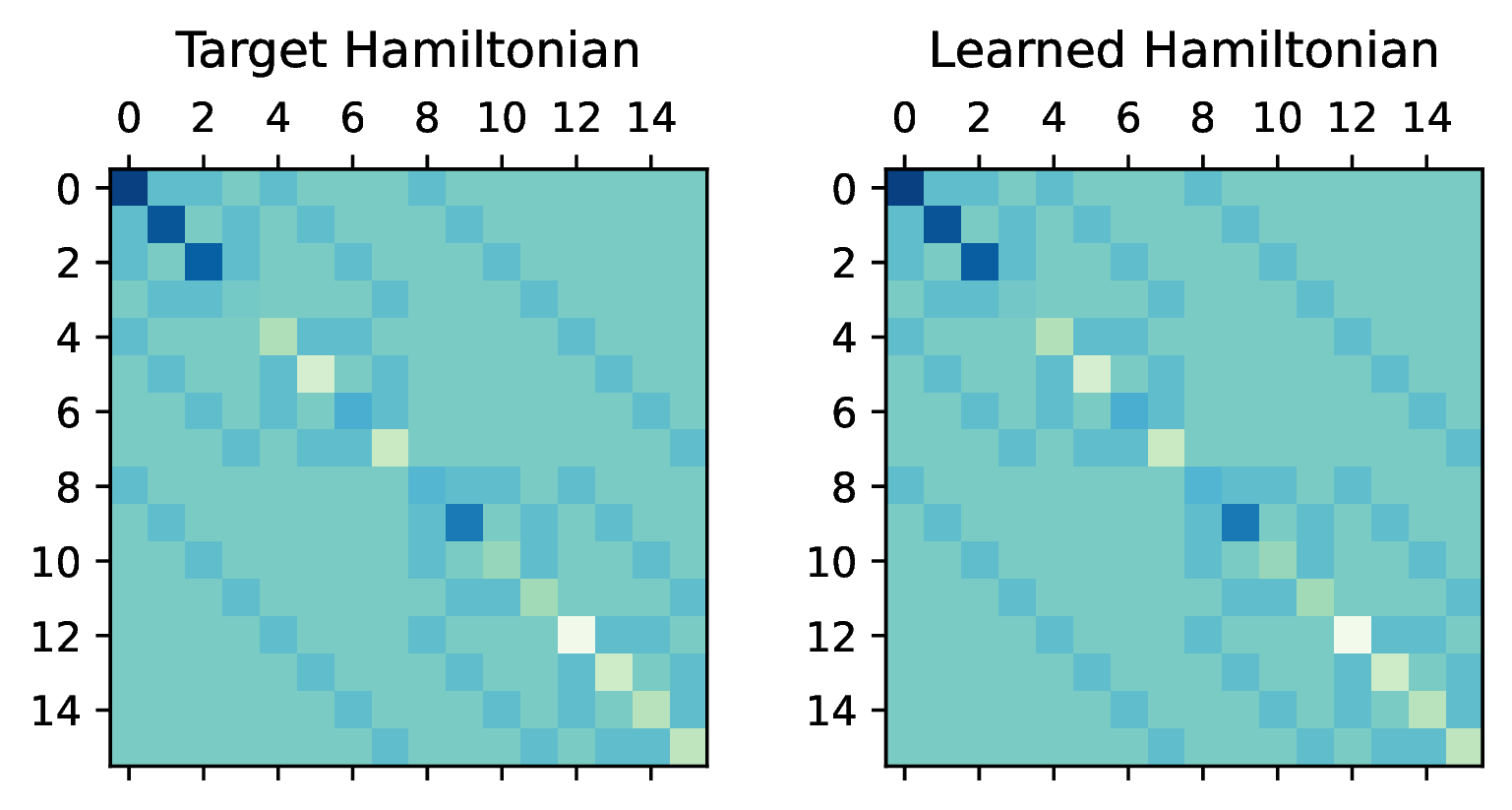}}
    \subfigure[Target and learned Hamiltonian for Sample 3]{\includegraphics[width=0.49\textwidth]{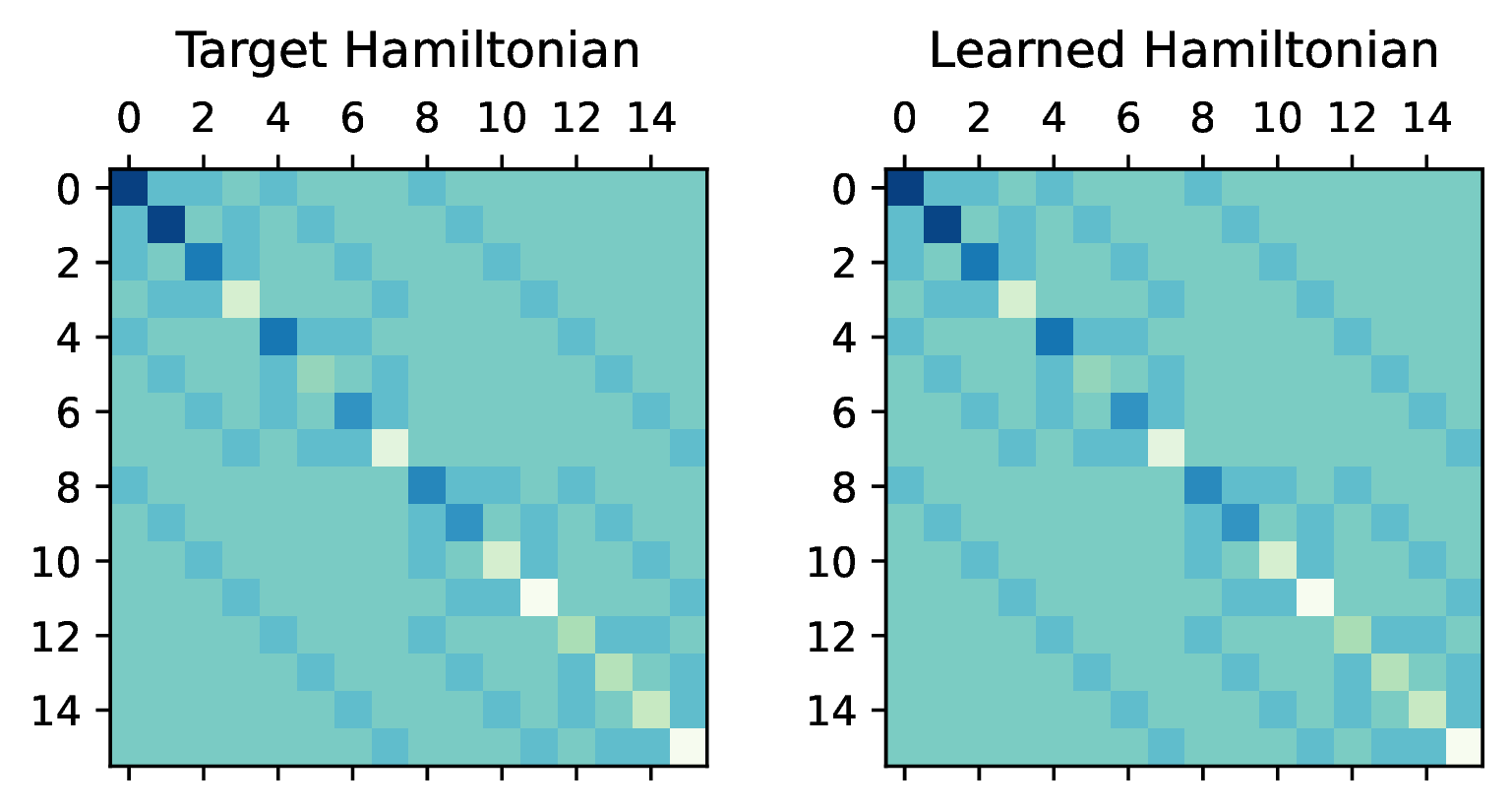}}
    \subfigure[Target and learned Hamiltonian for Sample 4]{\includegraphics[width=0.49\textwidth]{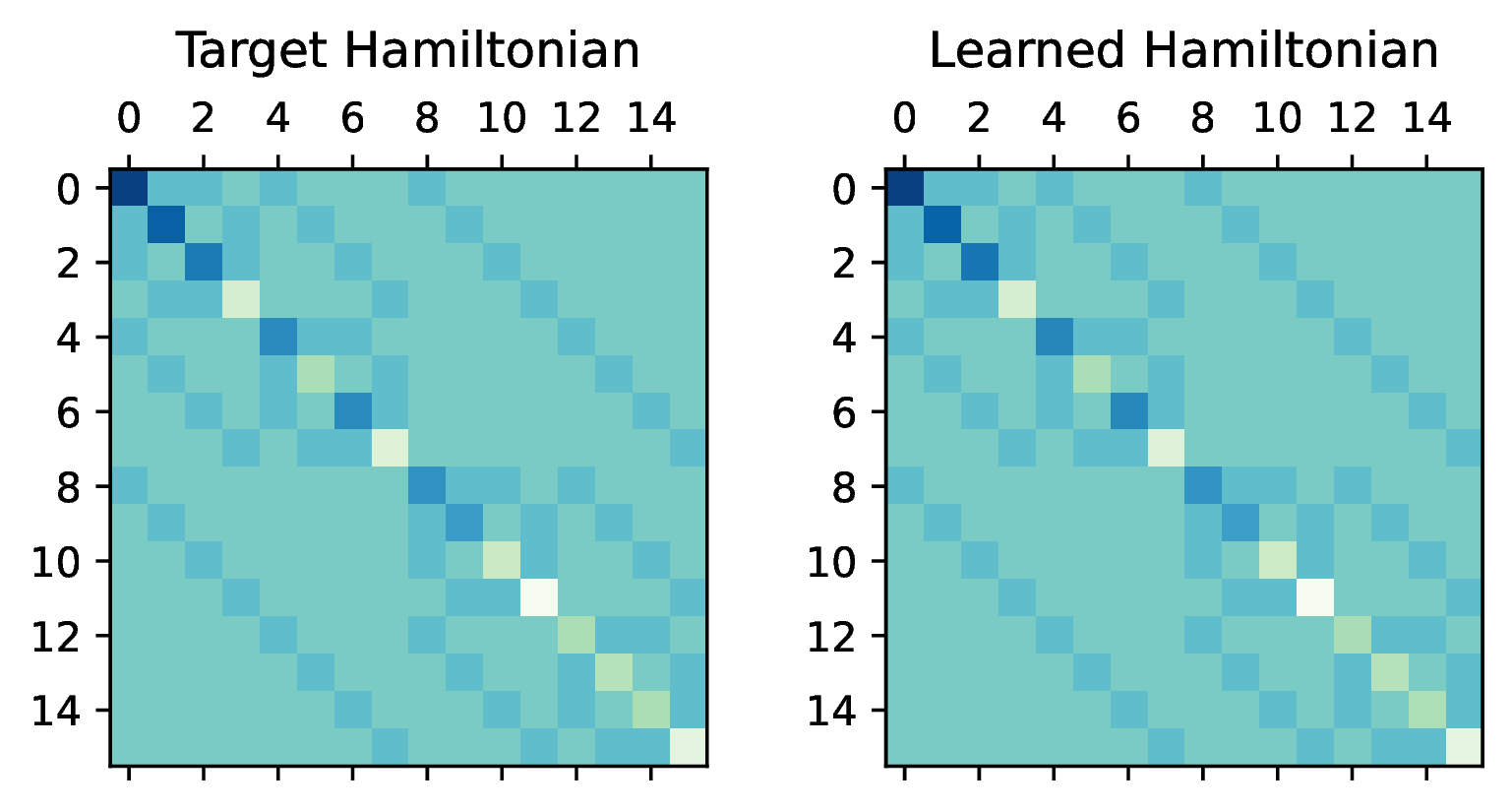}}
    \subfigure[Target and learned Hamiltonian for Sample 5]{\includegraphics[width=0.49\textwidth]{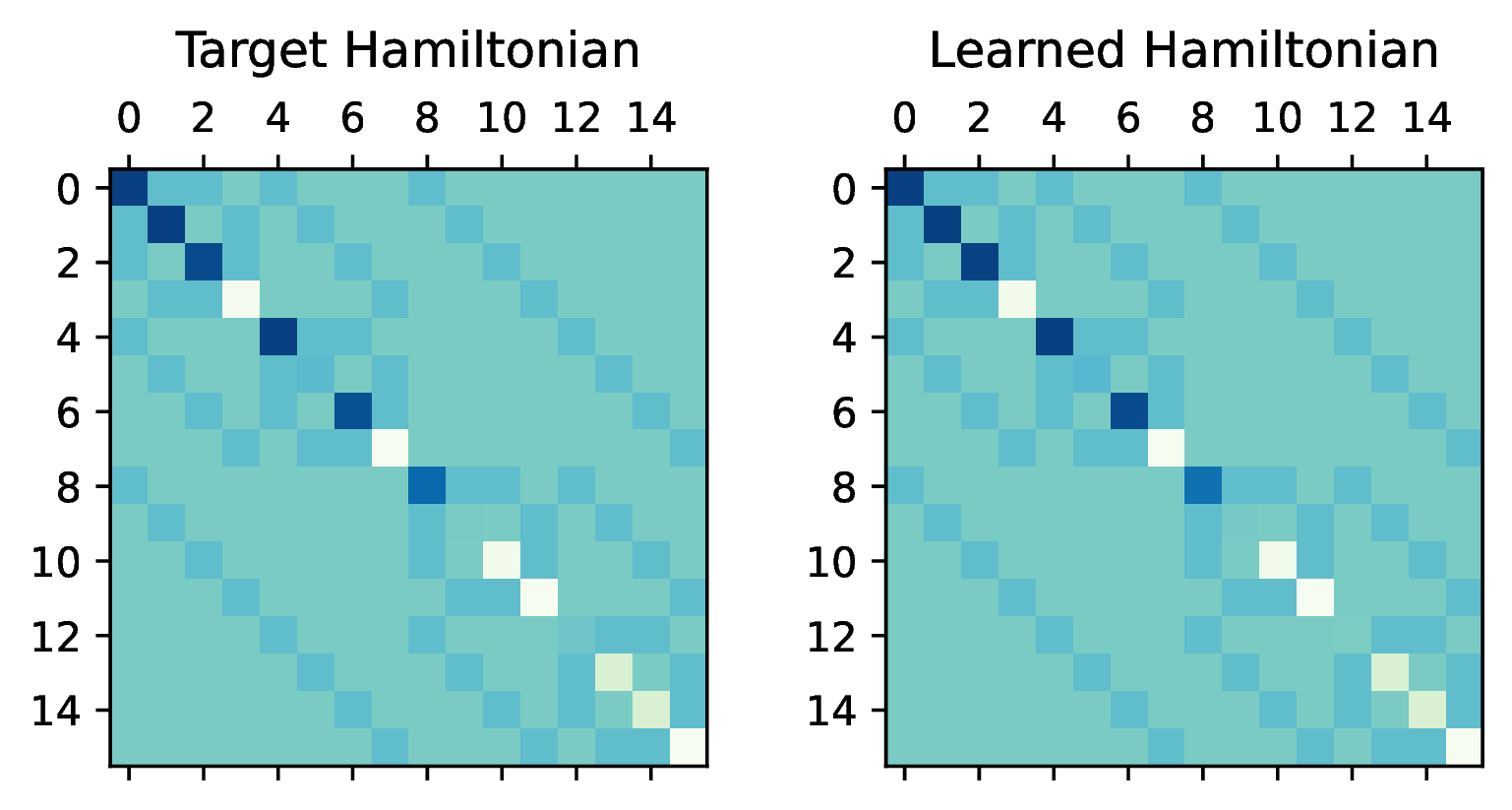}}
    \subfigure[Target and learned Hamiltonian for Sample 6]{\includegraphics[width=0.49\textwidth]{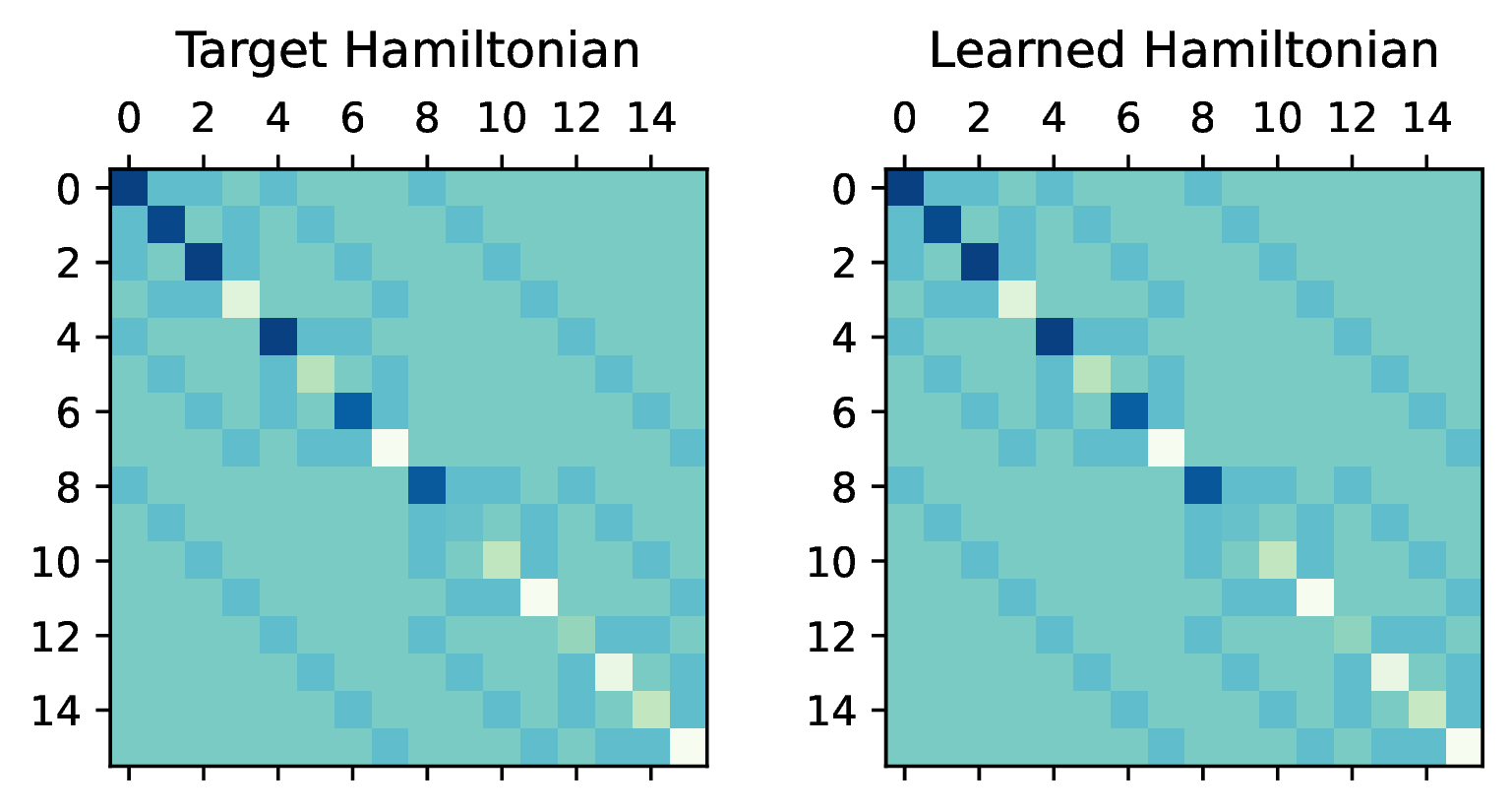}}  
    \caption{A visual representation of the Hamiltonian for all tested Iris samples}
    \label{fig:hamiltonians_iris}
\end{figure}

To further evaluate the model, we used the predicted nodes and edges to reconstruct the Hamiltonian. These reconstructed Hamiltonians were then compared to the actual (target) Hamiltonians of the system. Figure \ref{fig:hamiltonians_iris} presents the target Hamiltonian and the corresponding learned Hamiltonian for all six Iris samples. The structure of the learned Hamiltonians closely resembles that of the target Hamiltonians, demonstrating that QGRNN effectively captures the underlying relationships in the graph structure. While the Hamiltonian itself is not strictly related in our classical graph setting, its structural similarity to the target Hamiltonian provides a useful validation metric.

\subsubsection{Results on MNIST}
For the MNIST dataset, each image was transformed into six principal components using PCA. The goal is to assess how well QGRNN predicts the components that were embedded into a 6-node graph. To evaluate the accuracy, we apply the same evaluation metrics used for the Iris dataset, namely: MSE, RMSE, MAE, and Cosine Similarity. Additionally, we analyze the cost function convergence and compare the reconstructed Hamiltonians derived from the predicted node and edge values to the target Hamiltonians. 
Table \ref{tab:MNIST_predictions} presents a comparison between the actual PCA features of MNIST images and the corresponding features predicted by QGRNN. The predicted features closely match the actual features across all samples. 

\begin{table}[h!]
\centering
\caption{Comparison of Actual vs. Predicted Features for the MNIST Dataset}
\begin{tabular}{lll}
\toprule
\textbf{Sample} & \textbf{Actual Features} & \textbf{Predicted Features} \\ 
\midrule
      1 & (3.658, 1.960, 2.171, 2.878, 1.832, 2.447) & (3.700, 1.982, 2.235, 2.937, 1.852, 2.490) \\
      2 & (0.246, 1.631, 2.719, 2.507, 1.840, 1.784) & (0.234, 1.649, 2.770, 2.559, 1.854, 1.811) \\
      3 & (1.988, 1.905, 2.636, 3.849, 3.305, 3.116) & (2.010, 1.930, 2.687, 3.918, 3.330, 3.156) \\
      4 & (2.132, 1.014, 1.385, 1.824, 3.167, 1.592) & (2.160, 1.020, 1.423, 1.866, 3.208, 1.621) \\
      5 & (2.009, 3.527, 1.622, 3.890, 3.149, 2.916) & (2.030, 3.581, 1.662, 3.947, 3.192, 2.938) \\
      6 & (1.903, 1.864, 1.060, 1.980, 2.222, 2.249) & (1.925, 1.883, 1.084, 2.005, 2.250, 2.281) \\
      7 & (1.544, 2.466, 3.207, 2.953, 3.007, 0.709) & (1.574, 2.514, 3.260, 3.009, 3.044, 0.715) \\
      8 & (0.943, 3.087, 3.288, 1.458, 2.171, 3.015) & (0.963, 3.132, 3.347, 1.478, 2.189, 3.076) \\
      9 & (1.455, 1.890, 3.077, 0.902, 3.396, 2.787) & (1.457, 1.907, 3.132, 0.924, 3.447, 2.828) \\
     10 & (0.910, 3.480, 2.843, 1.436, 3.426, 2.281) & (0.903, 3.547, 2.894, 1.473, 3.467, 2.323) \\ \bottomrule
\end{tabular}
\label{tab:MNIST_predictions}
\end{table}

Table \ref{tab:MNIST_accuracy} provides a quantitative assessment of QGRNN’s prediction performance. MSE values range from 0.000642 to 0.002034, and RMSE values range from 0.025 to 0.045, confirming that the deviations between actual and predicted PCA features are minimal. MAE values are consistently low, with all samples showing MAE $\leq 0.0419$, indicating the absence of significant outliers. Cosine Similarity values are all above 0.99998, confirming that the predicted feature vectors maintain nearly perfect values. 
Despite the increased complexity of the MNIST dataset and its larger graph representation, QGRNN maintained high accuracy values. This indicates that the model scales effectively to higher-dimensional graphs without degradation in performance.

\begin{table}[h!]
\centering
\caption{Prediction Metrics for Each Sample in the MNIST Dataset}
\begin{tabular}{lllll}
\toprule
\textbf{Sample} & \textbf{MSE} & \textbf{RMSE} & \textbf{MAE} & \textbf{Cosine Similarity} \\ \hline
1 & 0.002034 & 0.045097 & 0.041873 & 0.999982 \\ 
2 & 0.001093 & 0.033057 & 0.028739 & 0.999984 \\
3 & 0.001782 & 0.042213 & 0.038665 & 0.999991 \\ 
4 & 0.001089 & 0.033003 & 0.030557 & 0.999986 \\ 
5 & 0.001778 & 0.042161 & 0.039702 & 0.999993 \\ 
6 & 0.000642 & 0.025338 & 0.025025 & 0.999997 \\ 
7 & 0.001738 & 0.041694 & 0.038049 & 0.999996 \\ 
8 & 0.001706 & 0.041298 & 0.036945 & 0.999993 \\ 
9 & 0.001353 & 0.036777 & 0.031289 & 0.999991 \\ 
10 & 0.001958 & 0.044249 & 0.040487 & 0.999988 \\ 
\bottomrule
\end{tabular}
\label{tab:MNIST_accuracy}
\end{table}

\begin{figure}[H]
    \centering
    \subfigure[Cost results for Sample 1]{\includegraphics[width=0.3\textwidth]{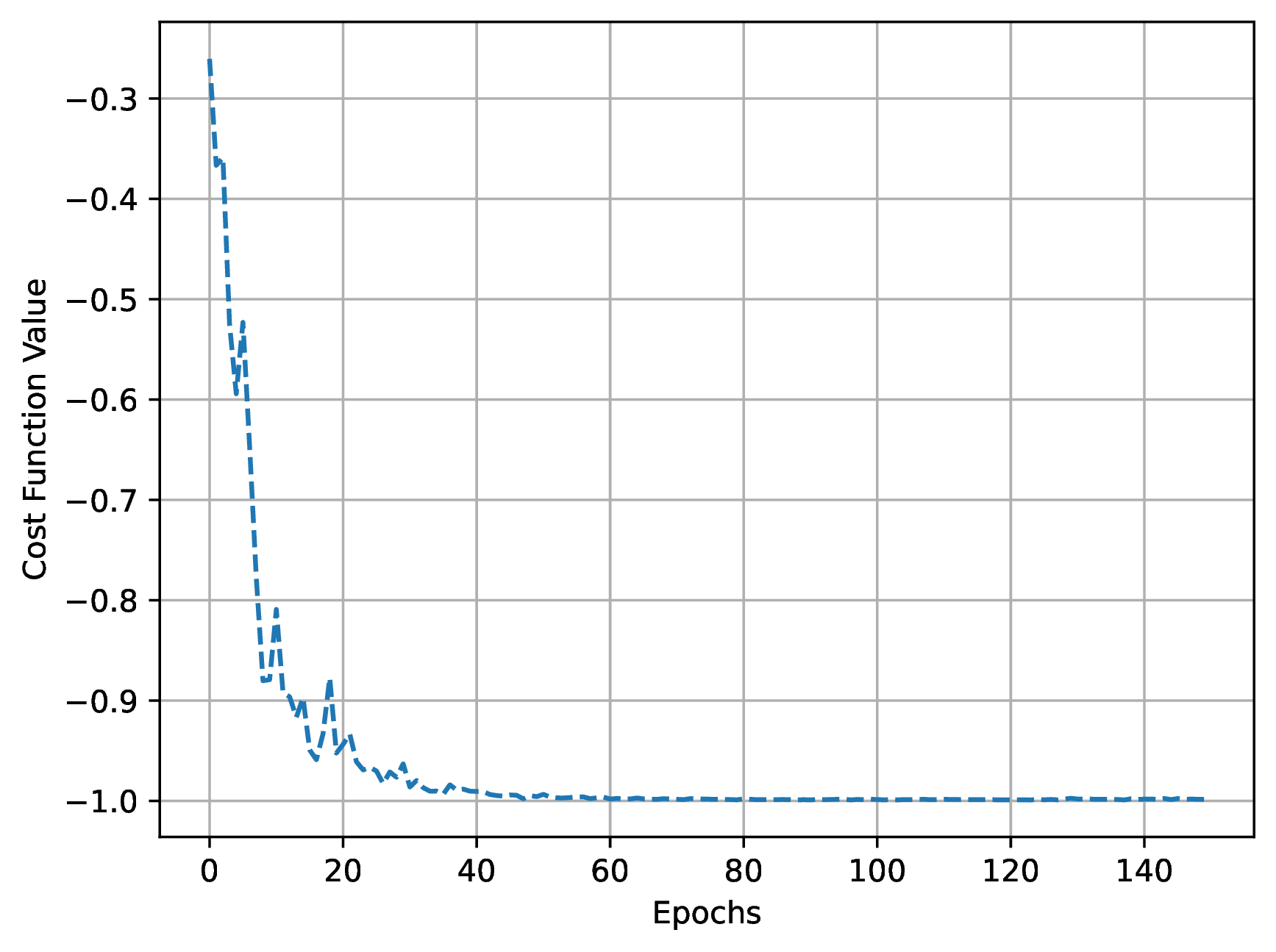}}
    \subfigure[Cost results for Sample 2]{\includegraphics[width=0.3\textwidth]{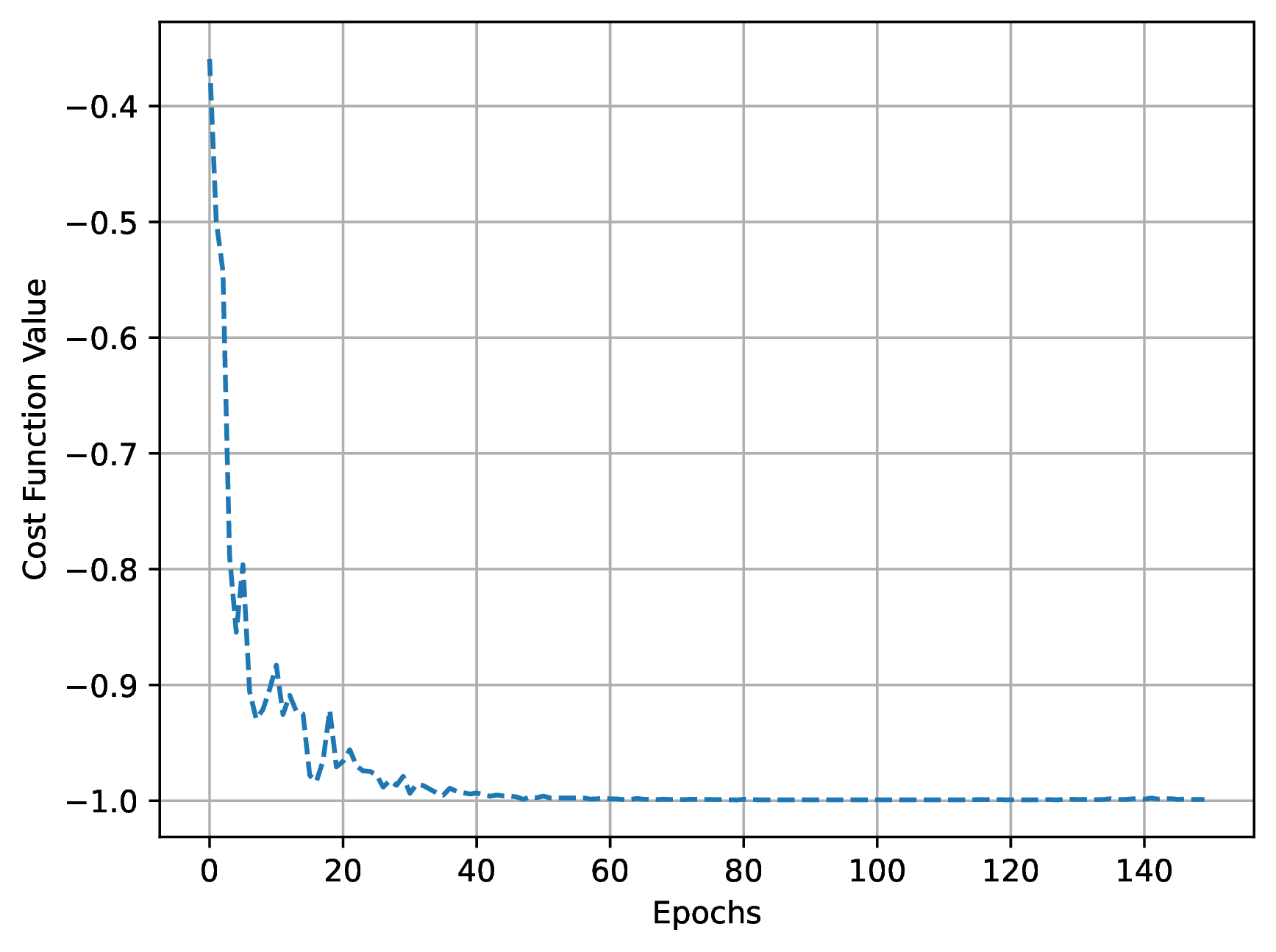}}
    \subfigure[Cost results for Sample 3]{\includegraphics[width=0.3\textwidth]{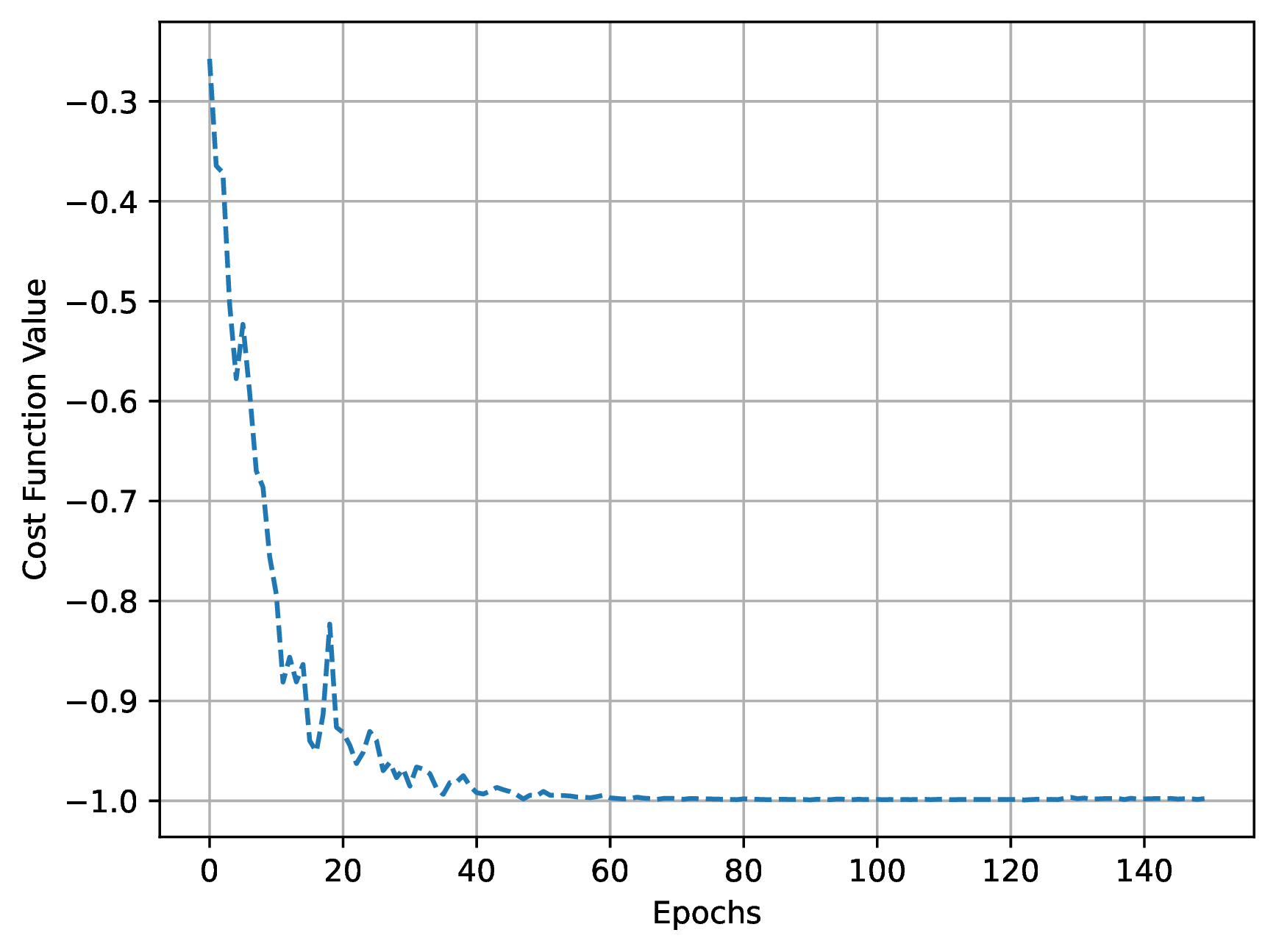}}
    \subfigure[Cost results for Sample 4]{\includegraphics[width=0.3\textwidth]{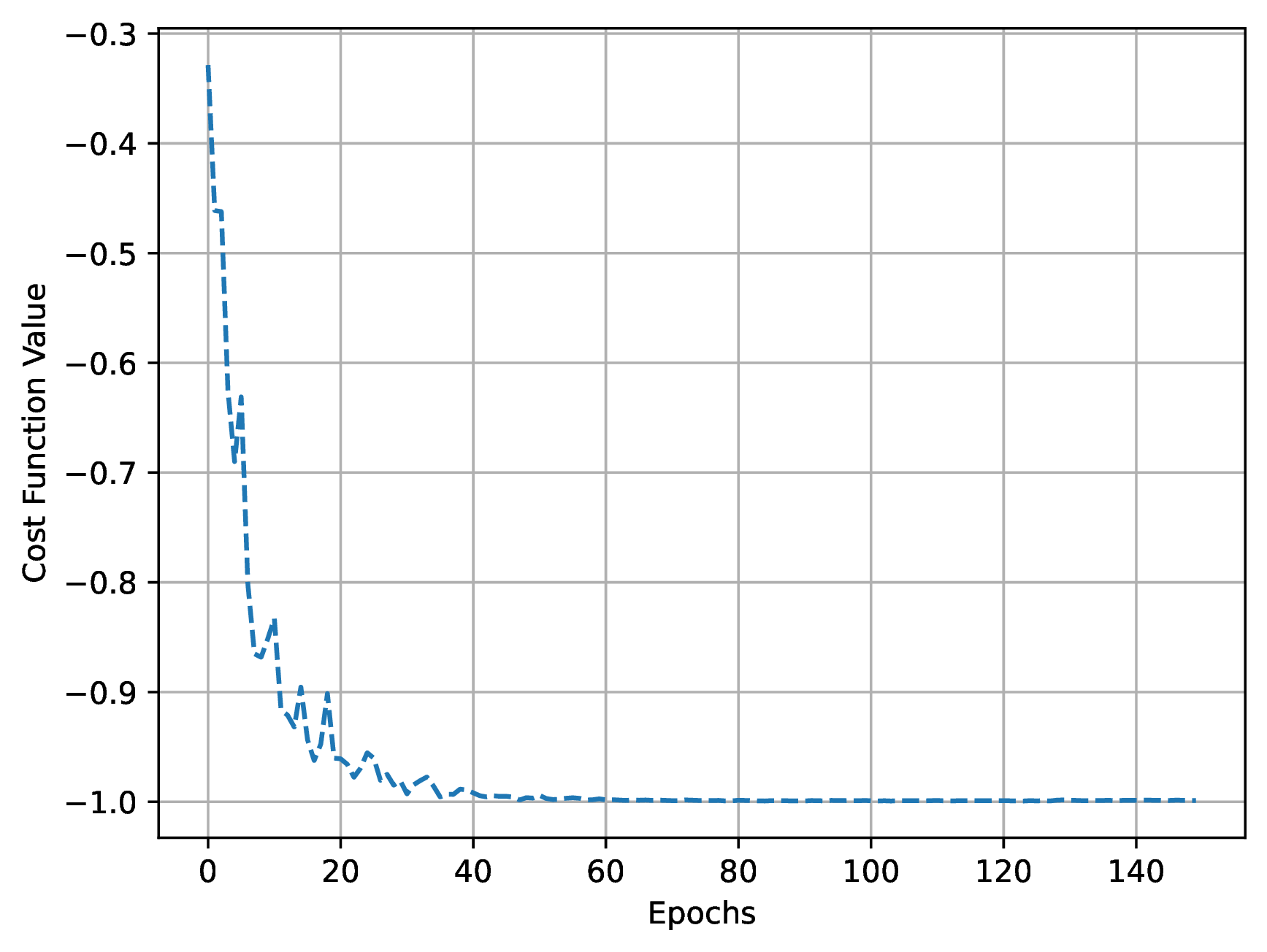}}
    \subfigure[Cost results for Sample 5]{\includegraphics[width=0.3\textwidth]{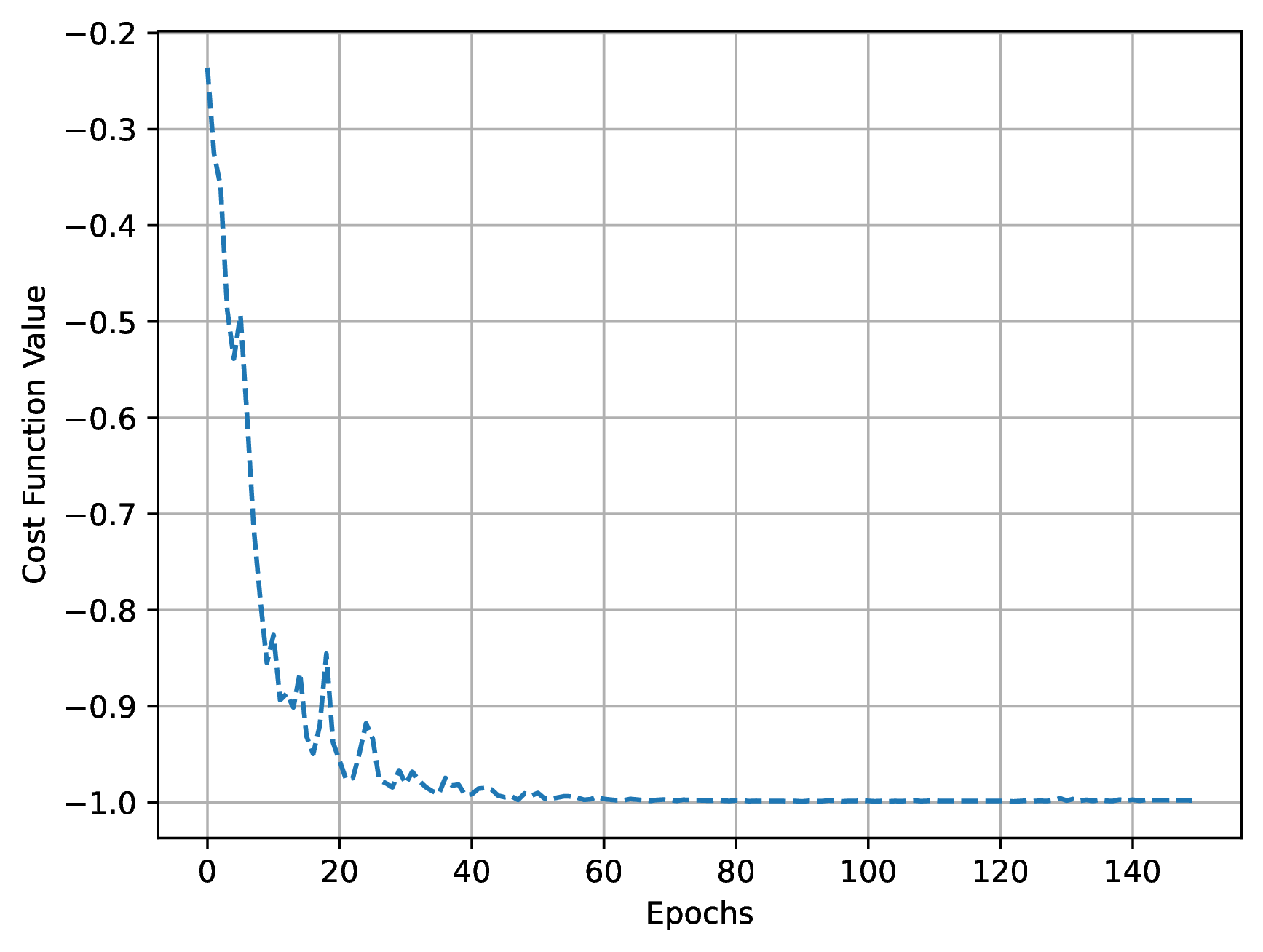}}
    \subfigure[Cost results for Sample 6]{\includegraphics[width=0.3\textwidth]{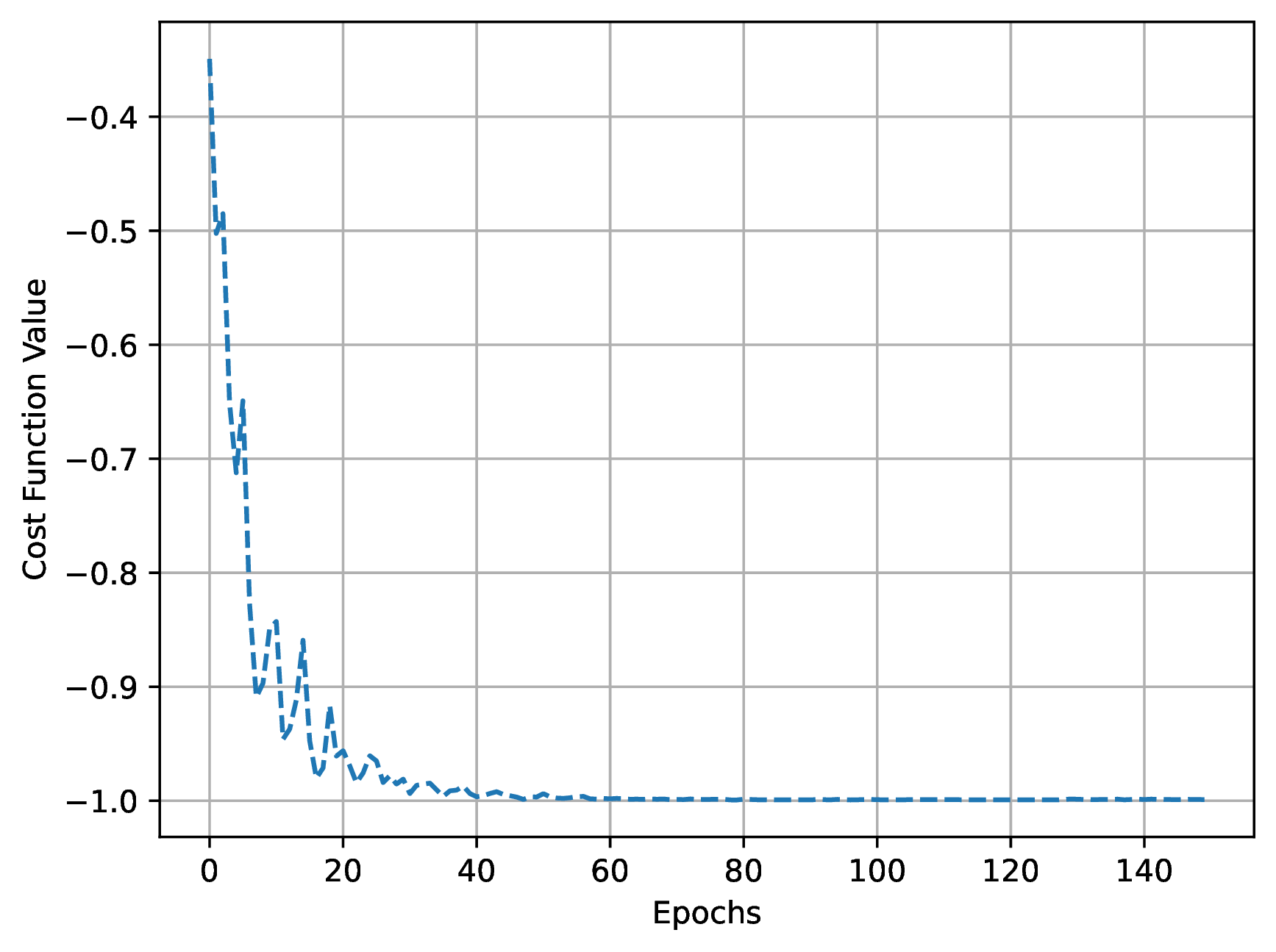}}
    \subfigure[Cost results for Sample 7]{\includegraphics[width=0.3\textwidth]{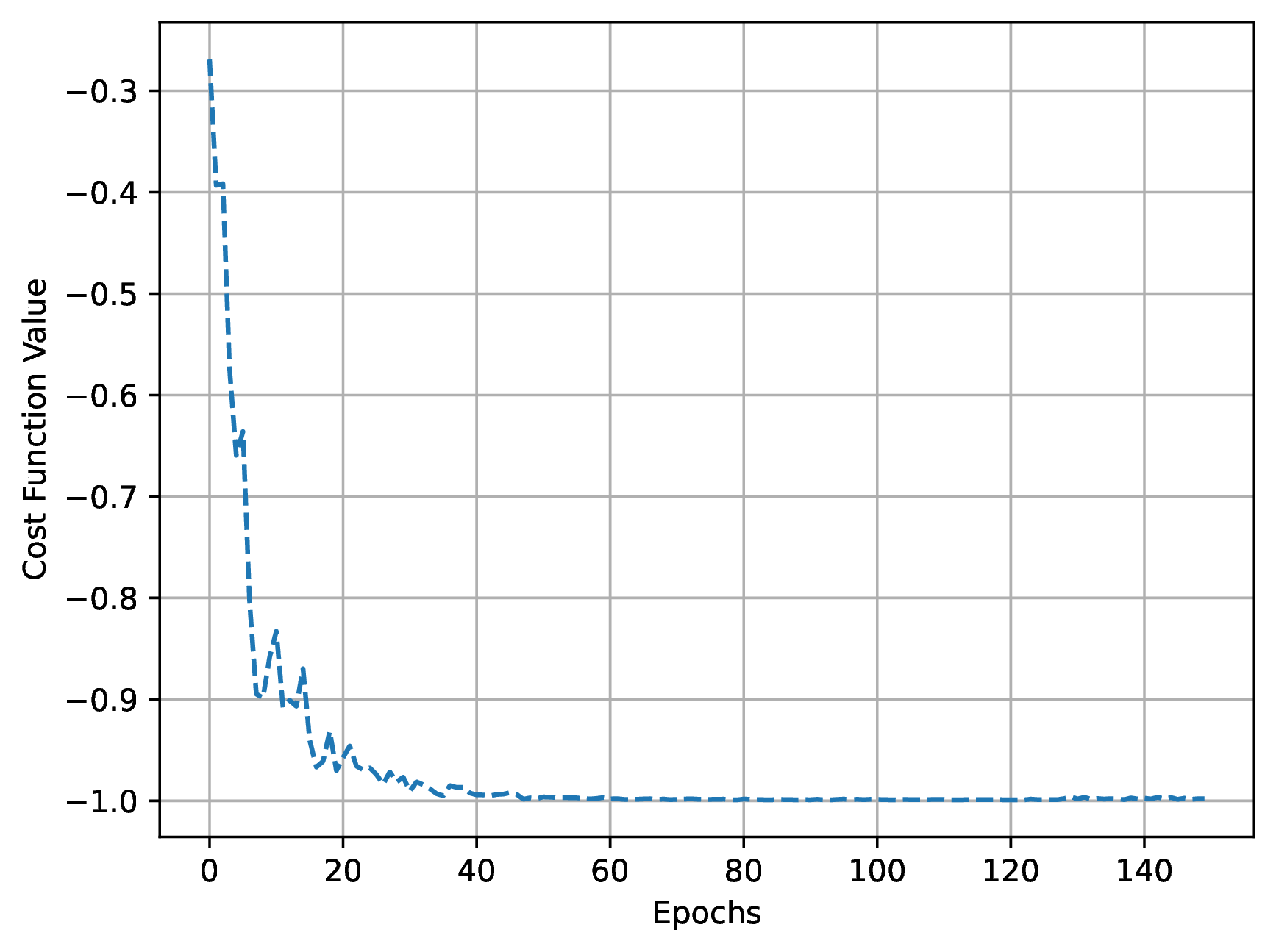}}  
    \subfigure[Cost results for Sample 8]{\includegraphics[width=0.3\textwidth]{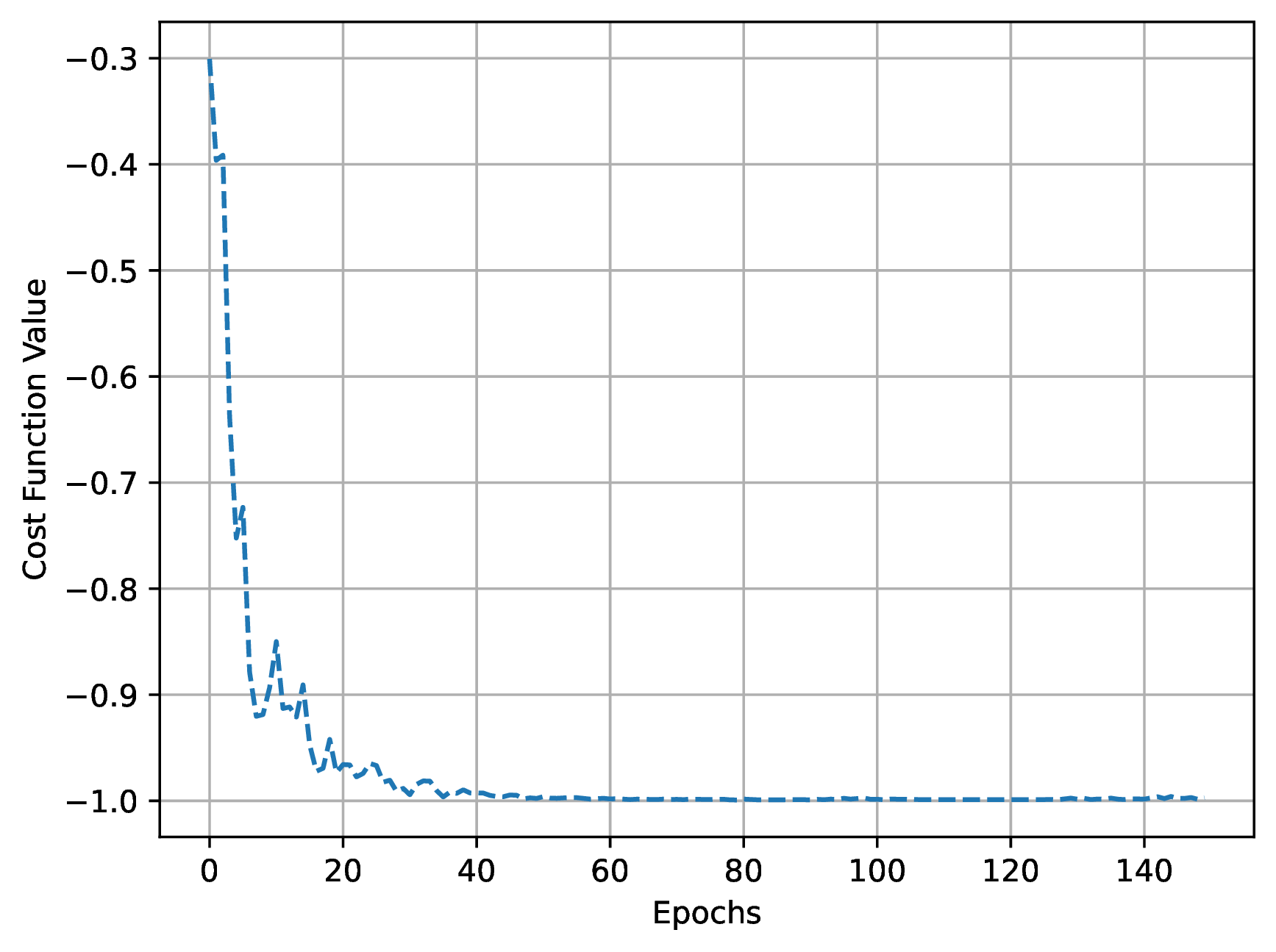}} 
    \subfigure[Cost results for Sample 9]{\includegraphics[width=0.3\textwidth]{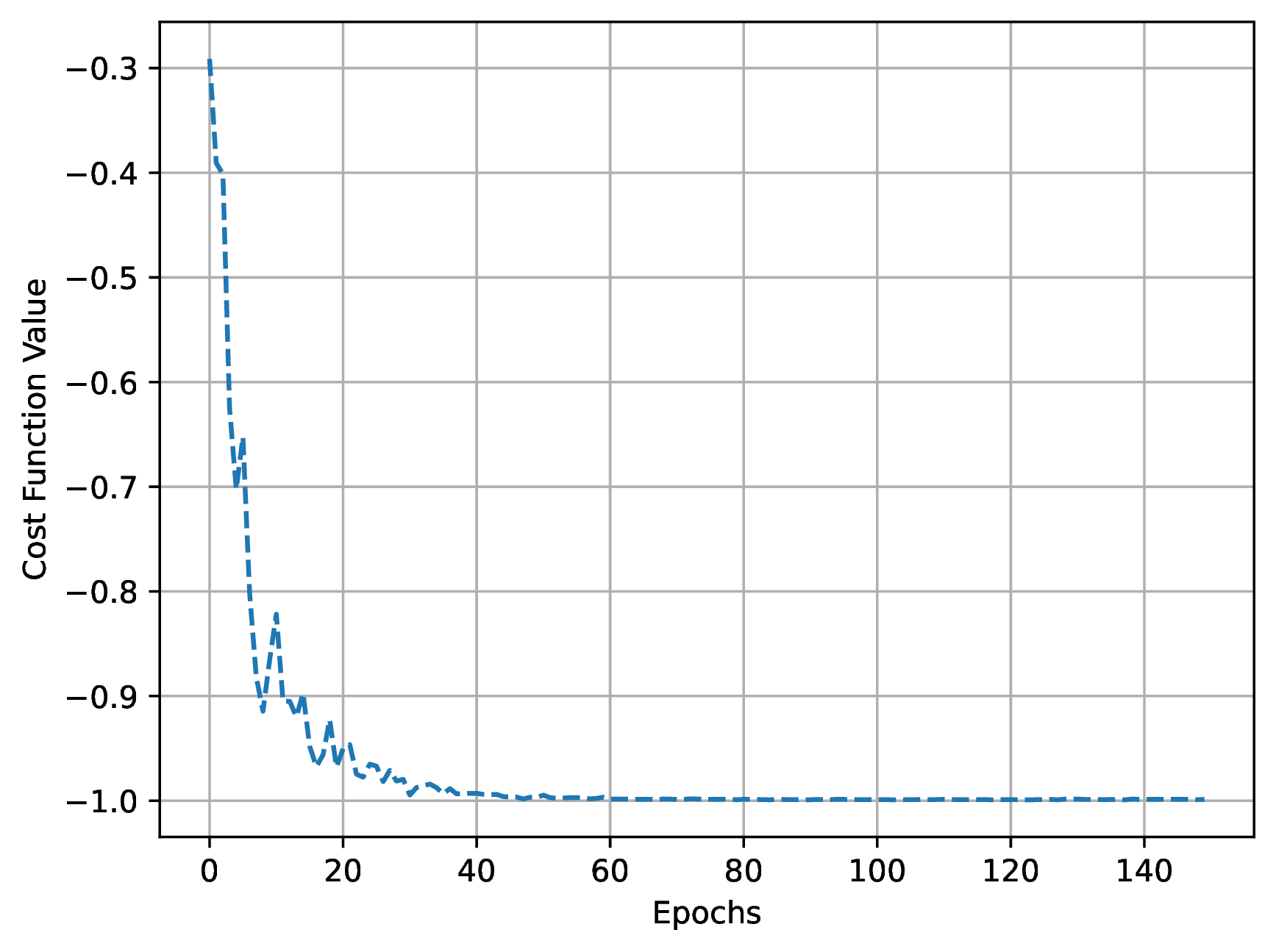}}  
    \subfigure[Cost results for Sample 10]{\includegraphics[width=0.3\textwidth]{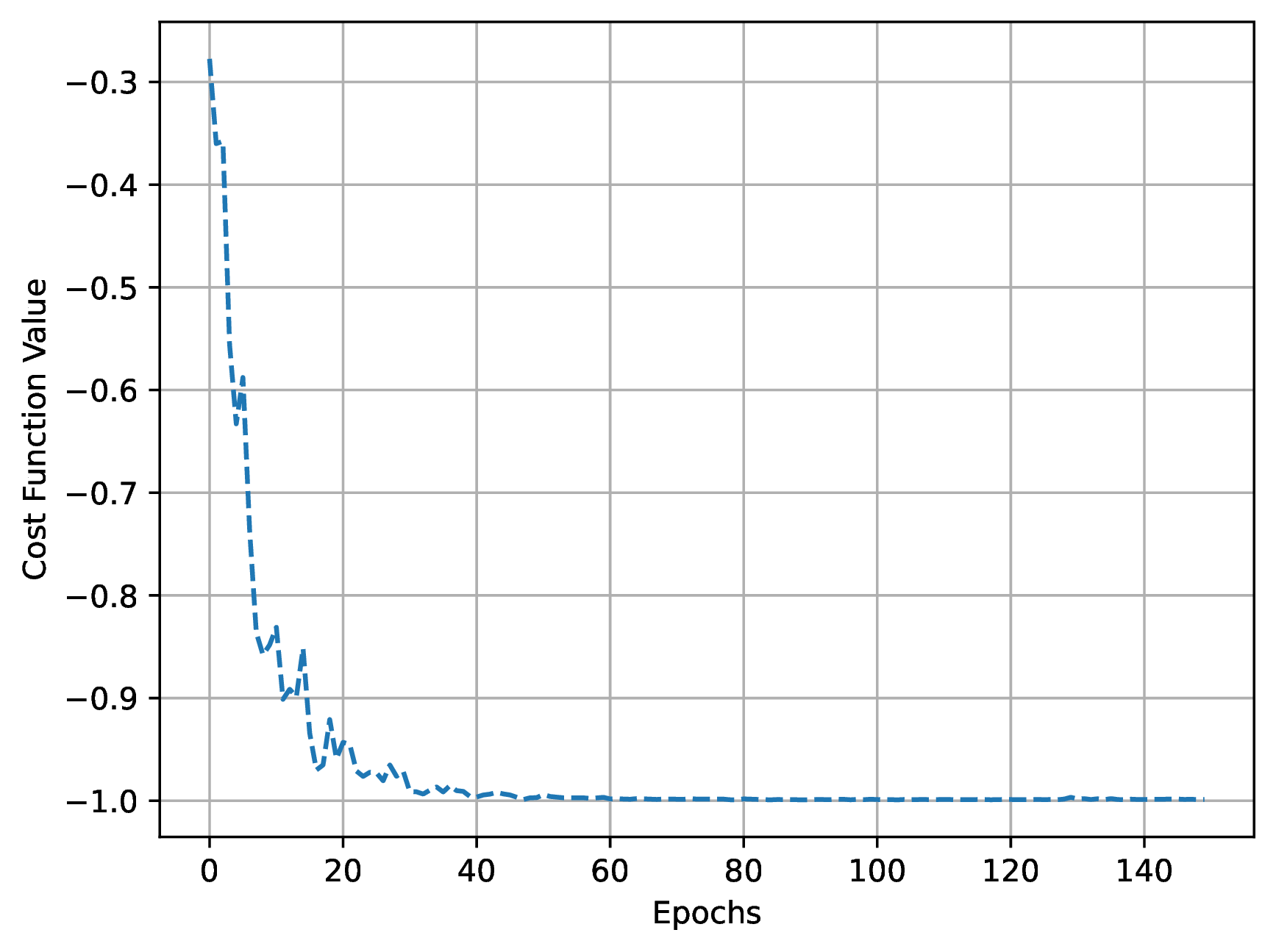}}  
    \caption{Cost function value for all tested MNIST samples}
    \label{fig:cost_function_MNIST}
\end{figure}

Figure \ref{fig:cost_function_MNIST} illustrates the cost function plots for all tested samples. The cost function value during training for the MNIST dataset showed a steady decrease over iterations, indicating successful learning by QGRNN. Despite the larger graph size and higher dimensionality of the data, the cost function converged smoothly across all samples, demonstrating the scalability of QGRNN.

\begin{figure}[H]
    \centering
    \subfigure[Target and learned Hamiltonian for Sample 1]{\includegraphics[width=0.49\textwidth]{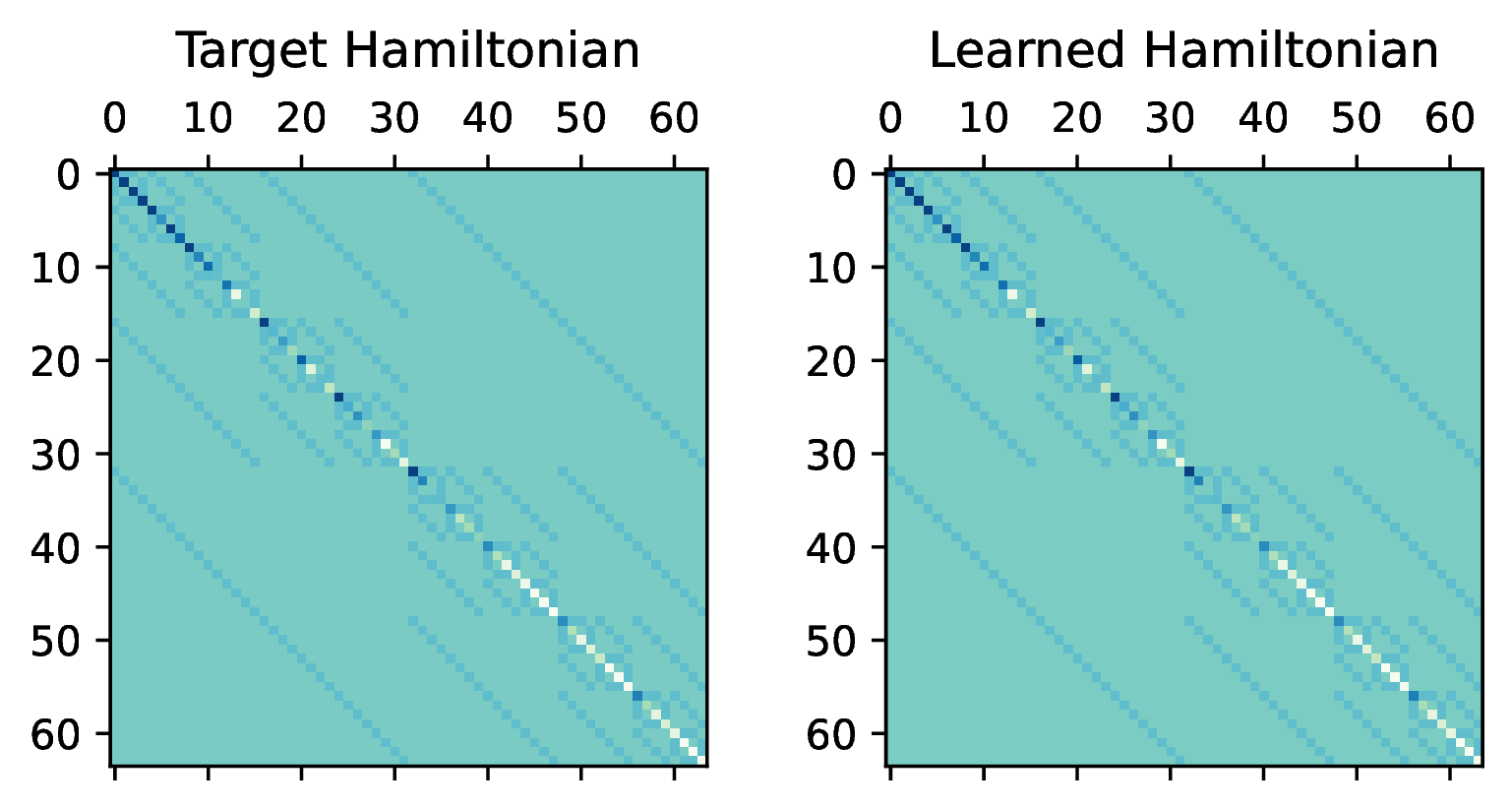}}
    \subfigure[Target and learned Hamiltonian for Sample 2]{\includegraphics[width=0.49\textwidth]{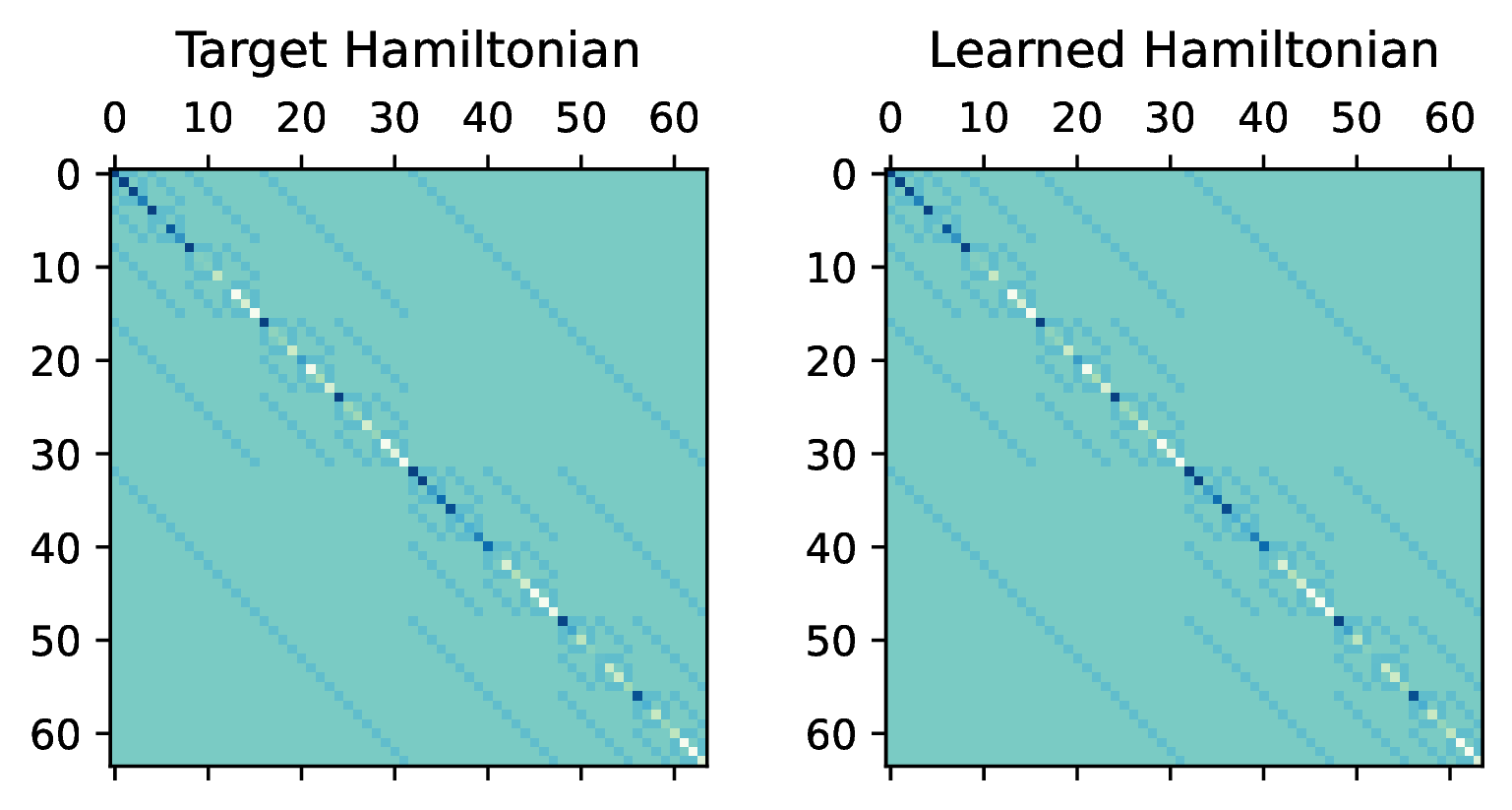}}
    \subfigure[Target and learned Hamiltonian for Sample 3]{\includegraphics[width=0.49\textwidth]{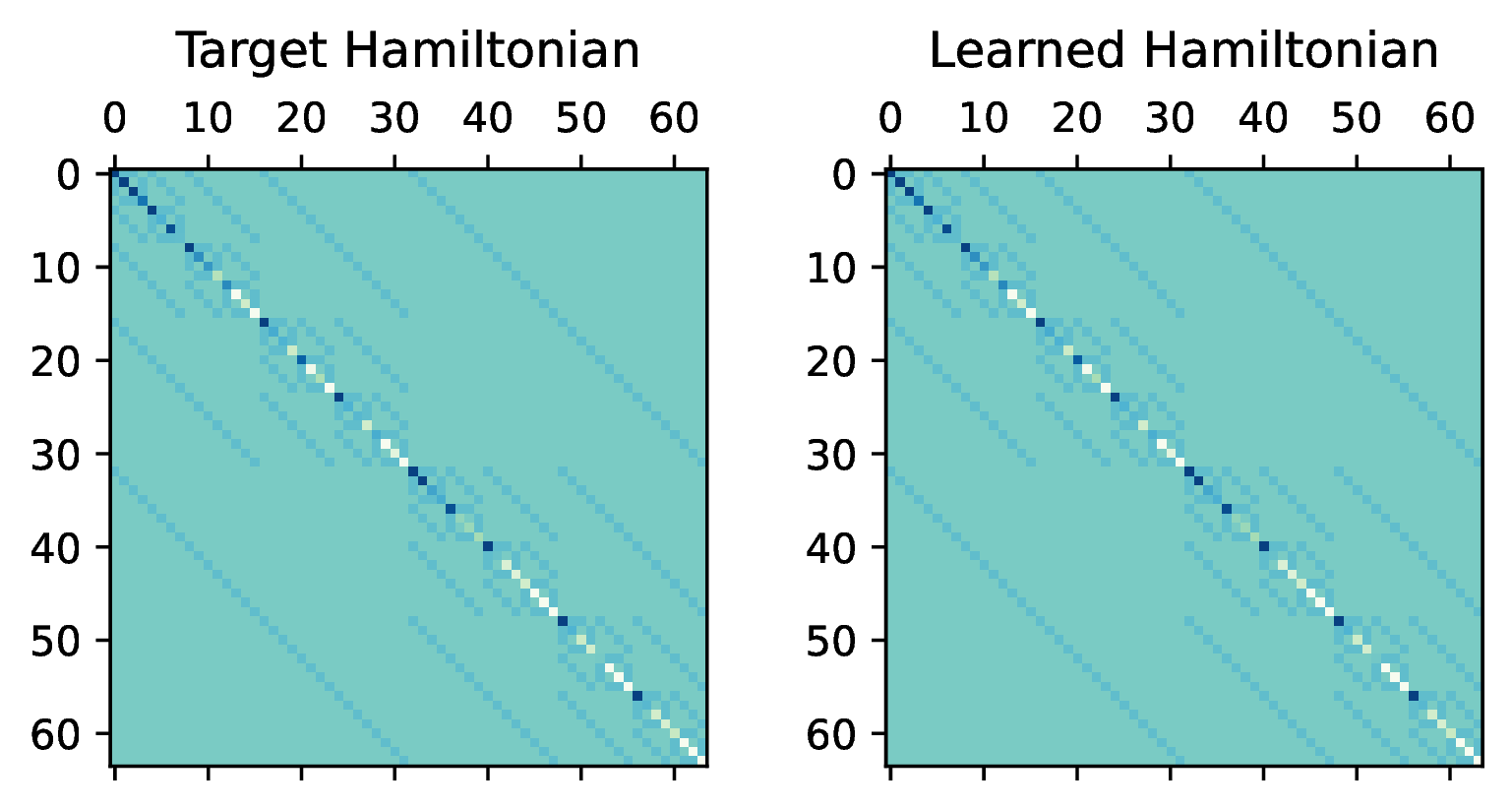}}
    \subfigure[Target and learned Hamiltonian for Sample 4]{\includegraphics[width=0.49\textwidth]{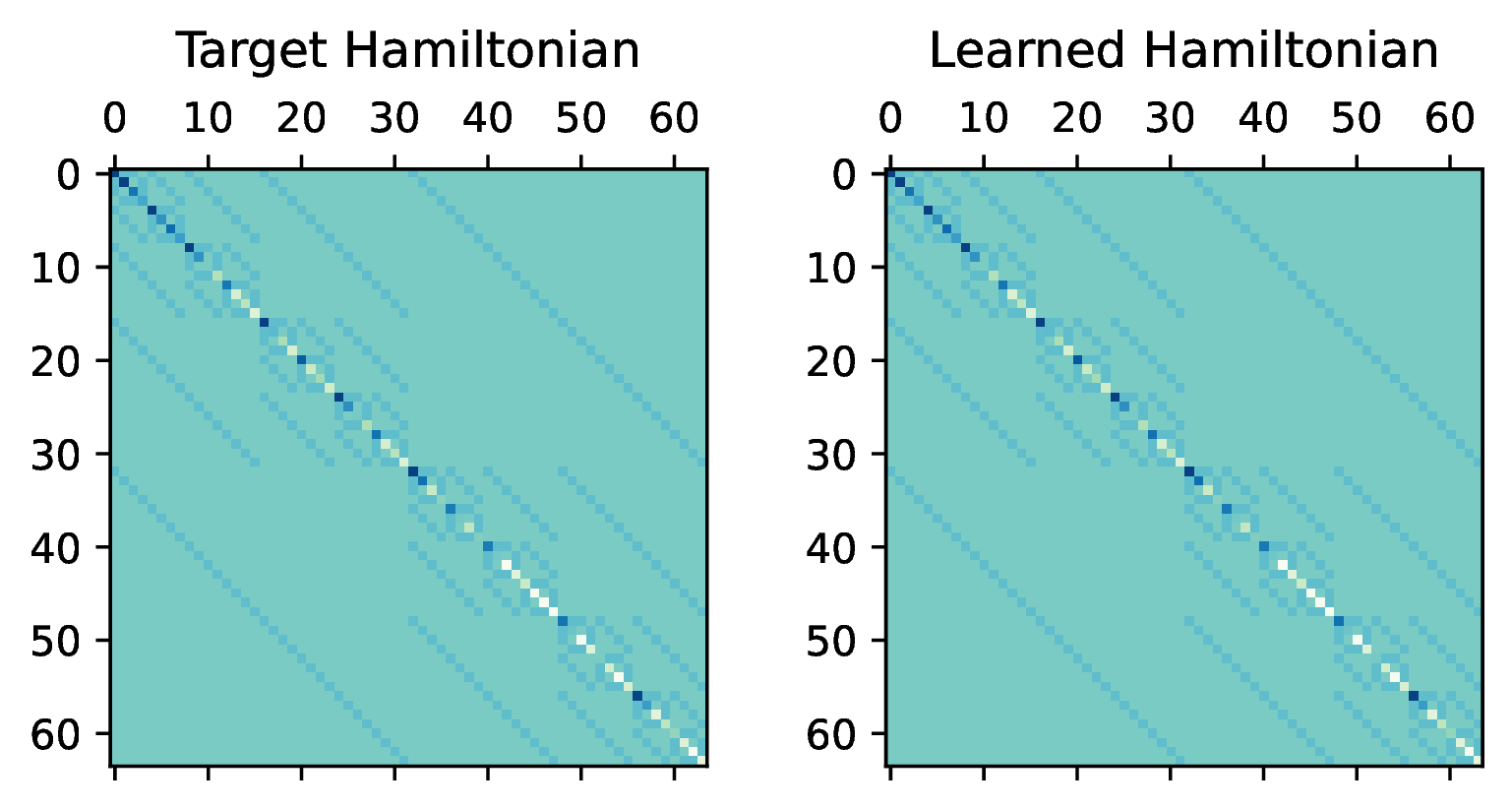}}
    \subfigure[Target and learned Hamiltonian for Sample 5]{\includegraphics[width=0.49\textwidth]{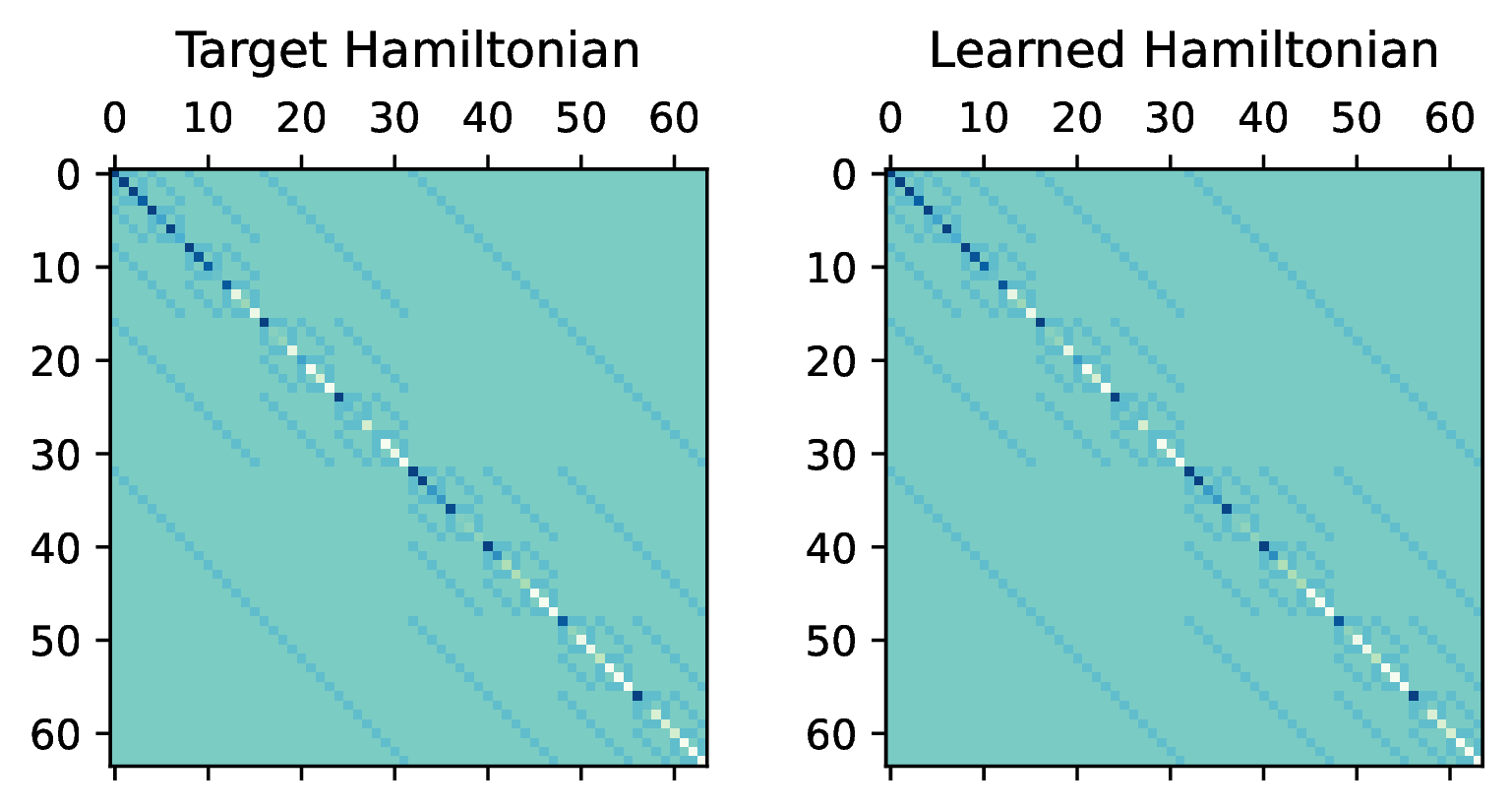}}
    \subfigure[Target and learned Hamiltonian for Sample 6]{\includegraphics[width=0.49\textwidth]{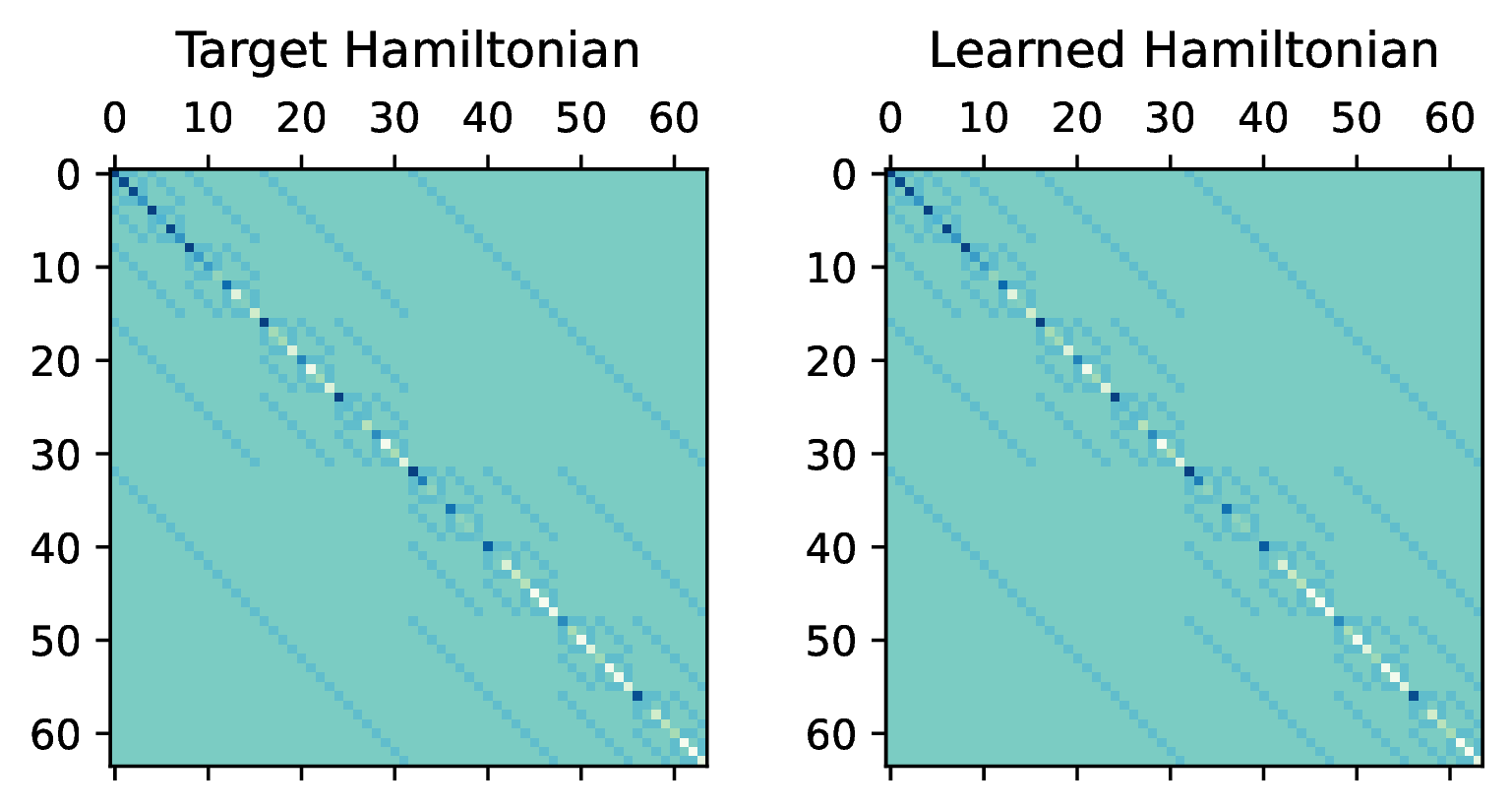}} 
    \subfigure[Target and learned Hamiltonian for Sample 7]{\includegraphics[width=0.49\textwidth]{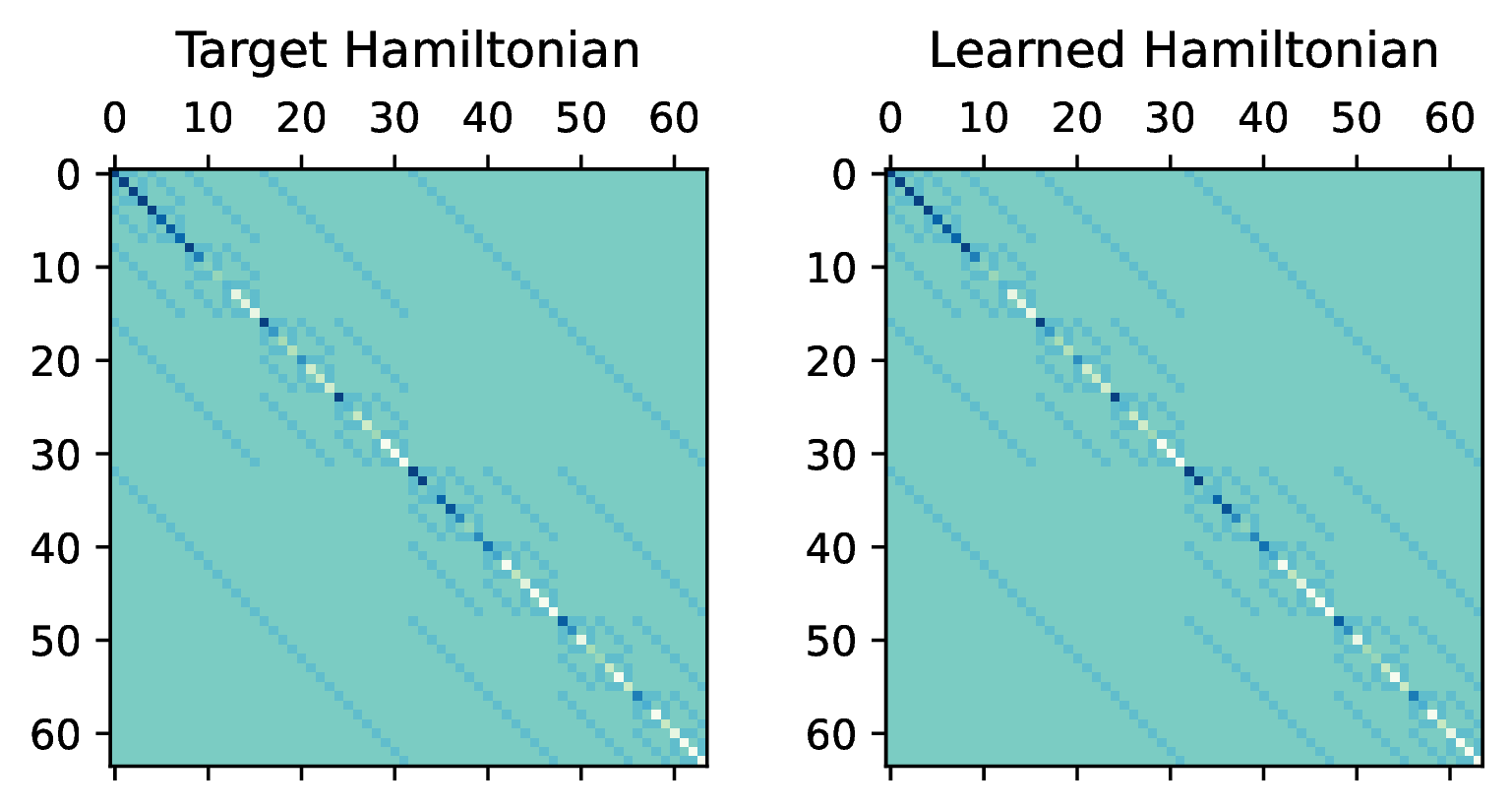}}  
    \subfigure[Target and learned Hamiltonian for Sample 8]{\includegraphics[width=0.49\textwidth]{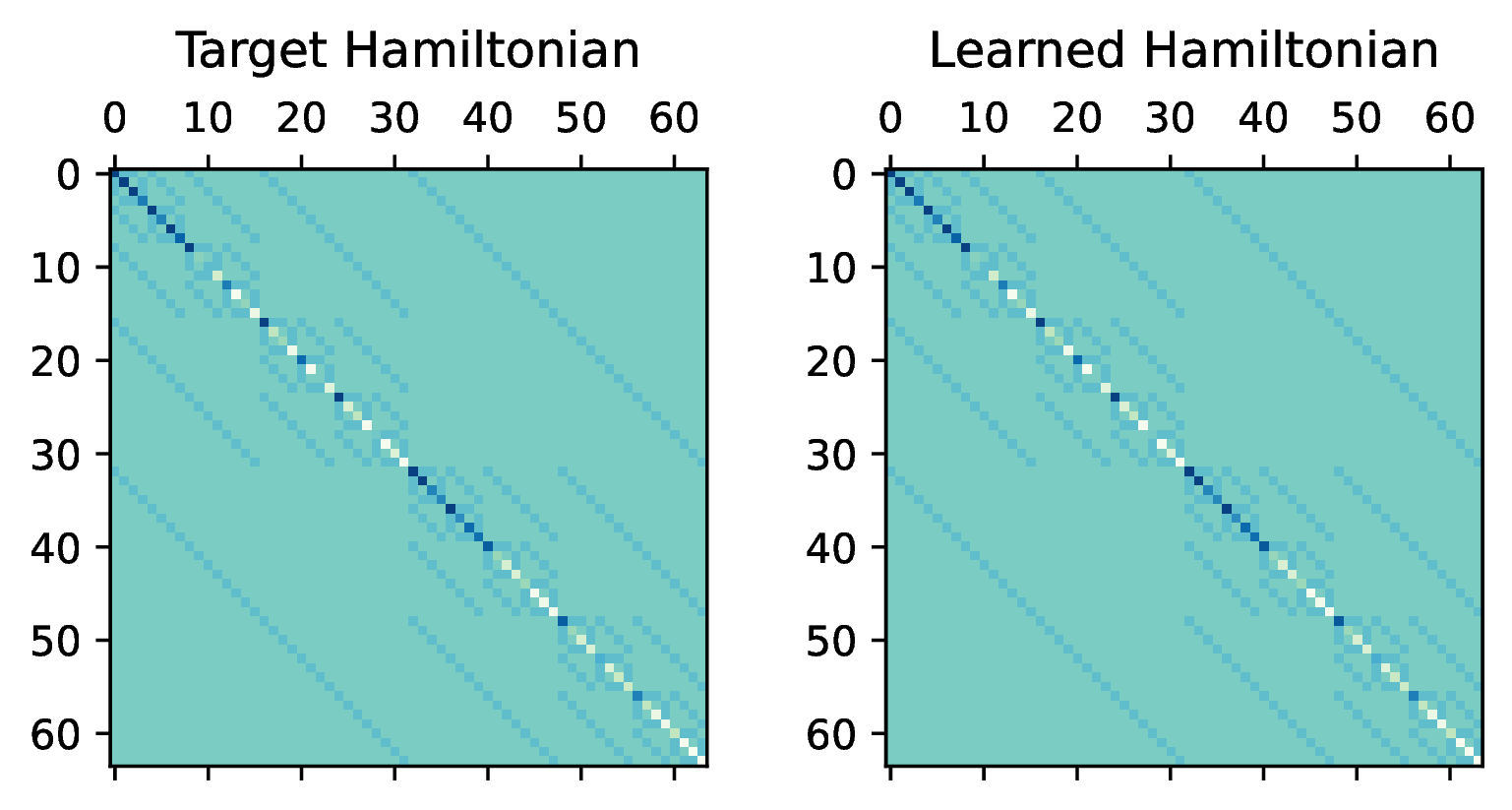}}  
    \subfigure[Target and learned Hamiltonian for Sample 9]{\includegraphics[width=0.49\textwidth]{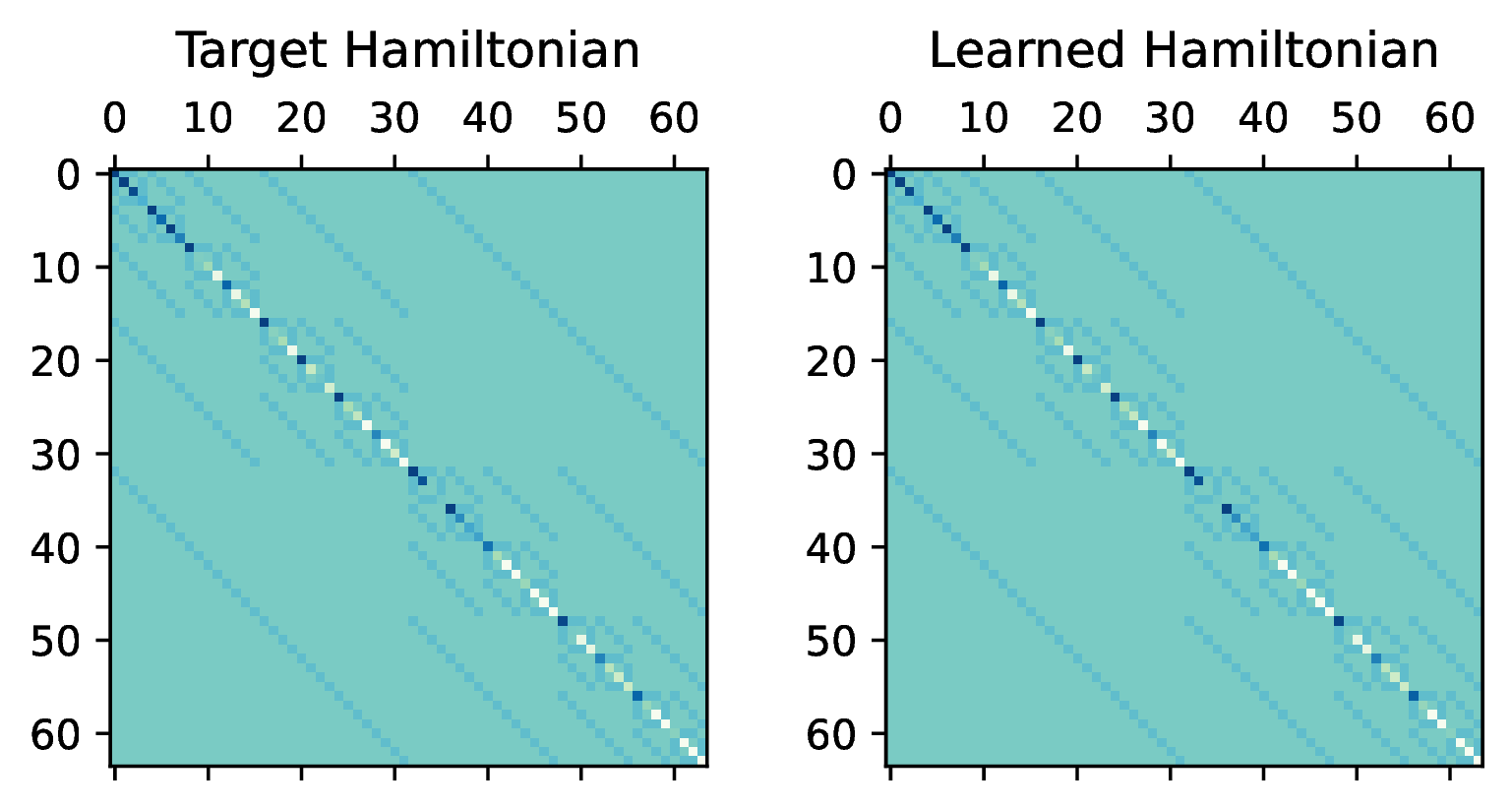}}  
    \subfigure[Target and learned Hamiltonian for Sample 10]{\includegraphics[width=0.49\textwidth]{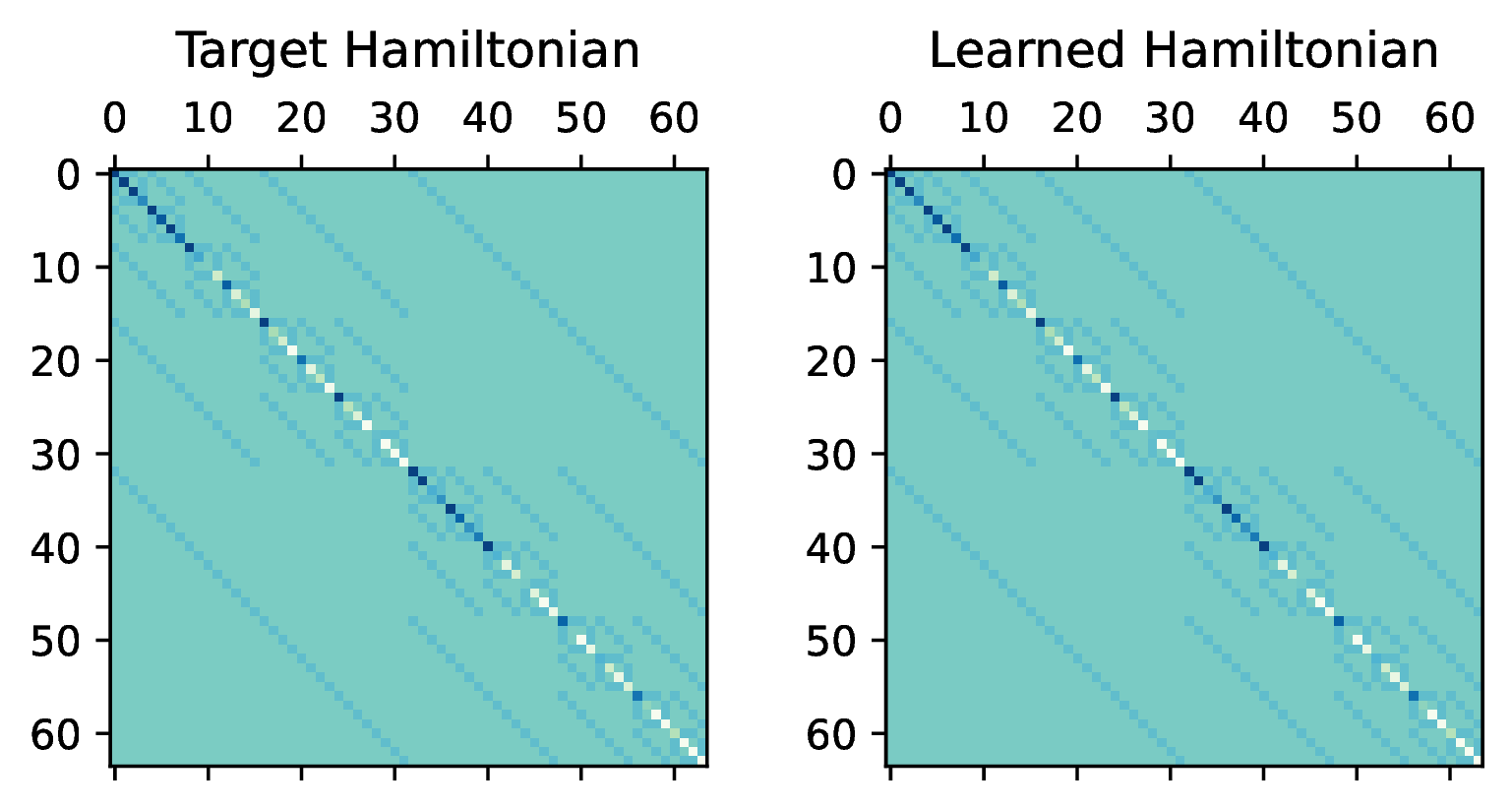}}  
    \caption{A visual representation of the Hamiltonian for all tested MNIST samples}
    \label{fig:hamiltonians_MNIST}
\end{figure}

Similarly to the Iris dataset, we also analyzed the Hamiltonian behavior for the MNIST dataset. Given the increased complexity of the MNIST graph, which required a higher number of qubits and nodes, we observed that QGRNN successfully approximated the Hamiltonian parameters with minimal deviation, where the learned Hamiltonian closely matched the target one. This confirms that QGRNN can effectively learn the underlying quantum graph structure even for larger datasets.

\subsection{Classification}
After QGRNN predicts the graph node features, the next step is to evaluate their effectiveness in classification tasks. To achieve this, we tested multiple classical machine learning models on the predicted features. The following classification models were used: Logistic Regression, Decision Tree, Random Forest, Naïve Bayes and Support Vector Machine (SVM). 

\subsubsection{Results on Iris}
We started by training classical classification models on the original Iris dataset. The classification accuracy of each model is shown in Table \ref{tab:iris_classification}. After training, we selected data points that were correctly classified and embedded them into a graph. We then applied our QGRNN-based methodology to reconstruct the feature values and classified these predicted features using the previously trained models. This approach allows us to evaluate whether the QGRNN-predicted features retain the class-distinguishing properties necessary for accurate classification.

\begin{table}[h!]
\centering
\caption{Classification accuracy of different models on the original Iris dataset}
\begin{tabular}{ll}
\toprule
\textbf{Classifier} & \textbf{Accuracy (\%)}\\ 
\midrule
Logistic Regression & 100 \\ 
Decision Tree & 100 \\ 
Random Forest & 100 \\ 
Naïve Bayes & 100 \\ 
Support Vector Machine (SVM) & 96.67 \\ 
\bottomrule
\end{tabular}
\label{tab:iris_classification}
\end{table}

Table \ref{tab:iris_qgrnn_classification} presents the classification results for six QGRNN-predicted samples. All tested classifiers achieved $100\%$  accuracy on the QGRNN-predicted features.
This indicates that QGRNN’s reconstructed features are nearly indistinguishable from the original dataset in terms of classification performance. Since the classifiers were trained on the actual Iris dataset and tested on QGRNN-predicted features, the perfect classification results confirm that QGRNN's feature reconstruction does not introduce any significant distortions.

\begin{table}[h]
\centering
\caption{Classification accuracy results for QGRNN-predicted features on Iris samples}
\begin{tabular}{cccccc}
\toprule
\textbf{Sample} & \textbf{Logistic Regression} & \textbf{Decision Tree} & \textbf{Random Forest} & \textbf{Naïve Bayes} & \textbf{SVM} \\ \midrule
1 & $100\%$ & $100\%$ & $100\%$ & $100\%$ & $100\%$ \\ 
2 & $100\%$ & $100\%$ & $100\%$ & $100\%$ & $100\%$ \\ 
3 & $100\%$ & $100\%$ & $100\%$ & $100\%$ & $100\%$ \\ 
4 & $100\%$ & $100\%$ & $100\%$ & $100\%$ & $100\%$ \\ 
5 & $100\%$ & $100\%$ & $100\%$ & $100\%$ & $100\%$ \\ 
6 & $100\%$ & $100\%$ & $100\%$ & $100\%$ & $100\%$ \\ 
\bottomrule
\end{tabular}
\label{tab:iris_qgrnn_classification}
\end{table}

\subsubsection{Results on MNIST}

Given that the Iris dataset is relatively simple (only 4 features per sample), the next step is to evaluate whether similar performance holds for MNIST, which involves a larger graph representation. We first trained classical classification models on the original PCA-reduced MNIST dataset. The classification accuracy of these models is presented in Table \ref{tab:MNIST_classification}.

\begin{table}[h!]
\centering
\caption{Classification Accuracy of Different Models on the Original PCA-reduced MNIST}
\begin{tabular}{ll}
\toprule
\textbf{Classifier} & \textbf{Accuracy (\%)}\\ 
\midrule
Logistic Regression & 73.05 \\ 
Decision Tree & 74.85 \\ 
Random Forest & 82.96 \\ 
Naïve Bayes & 69.71 \\ 
Support Vector Machine (SVM) & 76.33 \\ 
\bottomrule
\end{tabular}
\label{tab:MNIST_classification}
\end{table}

To ensure a controlled evaluation, we selected datapoints that were correctly classified in the original PCA-reduced MNIST and embedded them into a graph. Using our proposed methodology, QGRNN predicted the corresponding features, which were then fed into the previously trained models for classification. The classification results on QGRNN-predicted PCA-reduced MNIST features are presented in table \ref{tab:MNIST_qgrnn_classification}. The results indicate that QGRNN predicted features are nearly identical to the original PCA-reduced MNIST features in terms of classification performance.

\begin{table}[h]
\centering
\caption{Classification accuracy results for QGRNN-predicted features on PCA-reduced MNIST}
\begin{tabular}{llllll}
\toprule
\textbf{Sample} & \textbf{Logistic Regression} & \textbf{Decision Tree} & \textbf{Random Forest} & \textbf{Naïve Bayes} & \textbf{SVM} \\ \midrule
1 & $100\%$ & $100\%$ & $100\%$ & $100\%$ & $100\%$ \\ 
2 & $100\%$ & $100\%$ & $100\%$ & $100\%$ & $100\%$ \\ 
3 & $100\%$ & $100\%$ & $100\%$ & $100\%$ & $100\%$ \\ 
4 & $100\%$ & $100\%$ & $100\%$ & $100\%$ & $100\%$ \\ 
5 & $100\%$ & $100\%$ & $100\%$ & $100\%$ & $100\%$ \\ 
6 & $100\%$ & $100\%$ & $100\%$ & $100\%$ & $100\%$ \\ 
7 & $100\%$ & $100\%$ & $100\%$ & $100\%$ & $100\%$ \\ 
8 & $100\%$ & $100\%$ & $100\%$ & $100\%$ & $100\%$ \\  
9 & $100\%$ & $100\%$ & $100\%$ & $100\%$ & $100\%$ \\ 
10 & $100\%$ & $100\%$ & $100\%$ & $100\%$ & $100\%$ \\ 
\bottomrule
\end{tabular}
\label{tab:MNIST_qgrnn_classification}
\end{table}

The MNIST dataset required a larger graph representation compared to the Iris dataset, resulting in a higher number of qubits. Despite this increase in complexity, QGRNN successfully reconstructed the node features without degradation in the performance, which reinforces the robustness of QGRNN for processing high-dimensional data. 


\section{Information Hiding}\label{sec:applications}
The strong performance of QGRNN in reconstructing graph-based features and preserving classification accuracy demonstrates its potential for practical applications. The minimal difference between actual and predicted features, along with the scalability to larger graphs, indicates that QGRNN can be extended beyond standard classification tasks. Motivated by these promising results, in this section, we demonstrate the use of QRNN in constructing a robust information hiding scheme. 


Traditionally, information hiding has been achieved using classical computing techniques, such as steganography and cryptographic watermarking, which are applied to text, images, and audio \cite{agrawal2024perspective}. With the advent of quantum computing, new techniques for information hiding have emerged, leveraging the unique properties of quantum systems. In this work, we explore the use of QGRNN for information hiding, where the data is hidden in a graph, making retrieval possible only under specific quantum conditions. The hidden message can only be retrieved by reconstructing the graph parameters, ensuring that unauthorized parties cannot access the encoded data without knowledge of the initial and time-evolved states. This method offers enhanced security due to the no-cloning theorem, which prevents perfect duplication of quantum states, making unauthorized retrieval significantly more challenging.


The proposed information hiding method consists of two main steps: 
\begin{enumerate}
    \item \textbf{Encoding:} The message is embedded into a classical graph (in the nodes and edges). The system then evolves using its Hamiltonian, producing time-evolved states. It is crucial to keep the initial states used in the evolution process, as they serve as a reference for correctly extracting the hidden message. Without knowledge of these initial states, retrieving the encoded information becomes infeasible. The graph is then discarded, and only the time-evolved states and initial states are saved. 
    
    \item \textbf{Retrieval:} To extract the hidden message, one must recover the graph parameters using QGRNN. This is only possible if one has access to both the time-evolved states and the initial states. The recipient must reconstruct the graph parameters by optimizing QGRNN to minimize the loss function:
    \begin{equation}
 L(\beta, \delta) = -\frac{1}{N} \sum_{i=1}^N \left| \langle \psi(t_i) | U_H(\beta, \delta) | \psi_{initial} \rangle \right|^2   
\end{equation}

where $\beta$ corresponds to the initial Hamiltonian parameters 
\end{enumerate}

To evaluate the effectiveness of the proposed technique, we conduct experiments in which short messages are encoded into graphs of different sizes. There are various ways to achieve this. Our experiment involves converting words into numerical values and embedding them within a graph, either in the nodes, edges or both. In our experiment, we focus on converting words into numerical values and embedding them within a graph using only nodes, while ignoring edges. As a result, the sentence length directly determines the graph size.

\begin{figure}[H]
    \centering
    \includegraphics[width=0.7\linewidth]{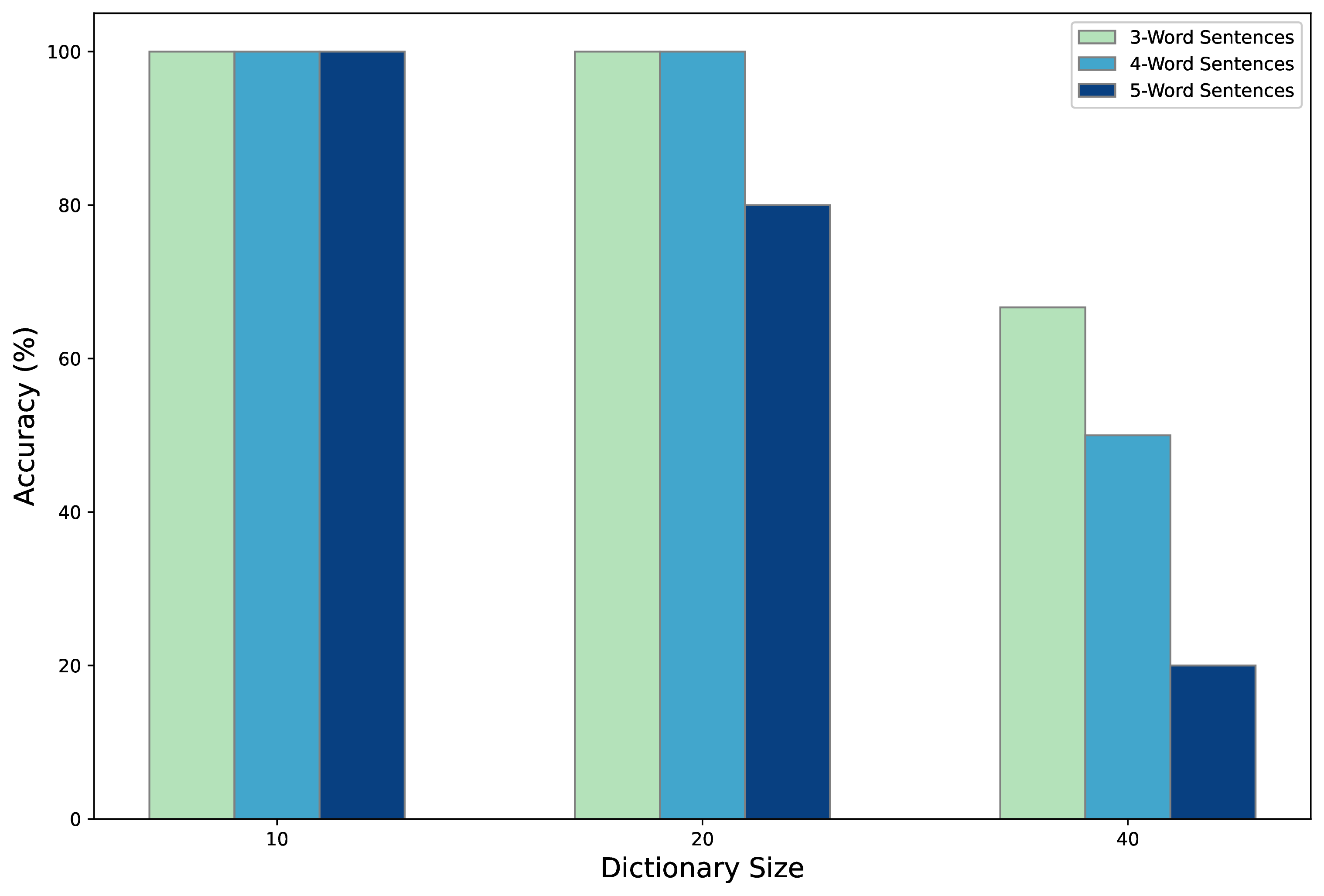}
    \caption{Accuracy Across Different Dictionary Sizes and Sentence Lengths}
    \label{fig:barchart}
\end{figure}

We start with a predefined dictionary of words, where each word in the dictionary is assigned a unique numerical value, and these values are spaced evenly within a given range. The spacing is calculated as:

\[
\text{Spacing} = \frac{\text{Number of Words} - 1}{\text{Total Range}}
\]

\noindent where the total range is taken from $[-4,5]$. For example, a 10-word dictionary results in a spacing of $1.0$, which means that each word is assigned values such as $[-4.0, -3.0, ..., 5.0]$. A 20-word dictionary results in a smaller spacing of 0.4737, and a 40-word dictionary has an even finer spacing of 0.2308. As the dictionary size increases, the spacing between assigned values decreases, making it harder to distinguish words numerically. With this setup, we conducted experiments using different dictionary sizes and sentence lengths to assess how well the encoded information can be retrieved. Figure \ref{fig:barchart} shows the accuracy for different dictionary sizes and sentence lengths. The accuracy was calculated using:
\[
\text{Accuracy} = \left( \frac{\text{Number of Correctly Retrieved Words}}{\text{Sentence Length}} \right) \times 100
\]

For 10-word and 20-word dictionaries, the accuracy was nearly perfect for all sentence lengths. For 40-word dictionaries, accuracy drops significantly, 3-word sentences still achieve a reasonable accuracy, but 4-word and 5-word sentences suffer more. 

\begin{figure}[H]
    \centering
    \subfigure[Dictionary = 10, 3-word]{\includegraphics[width=0.3\textwidth]{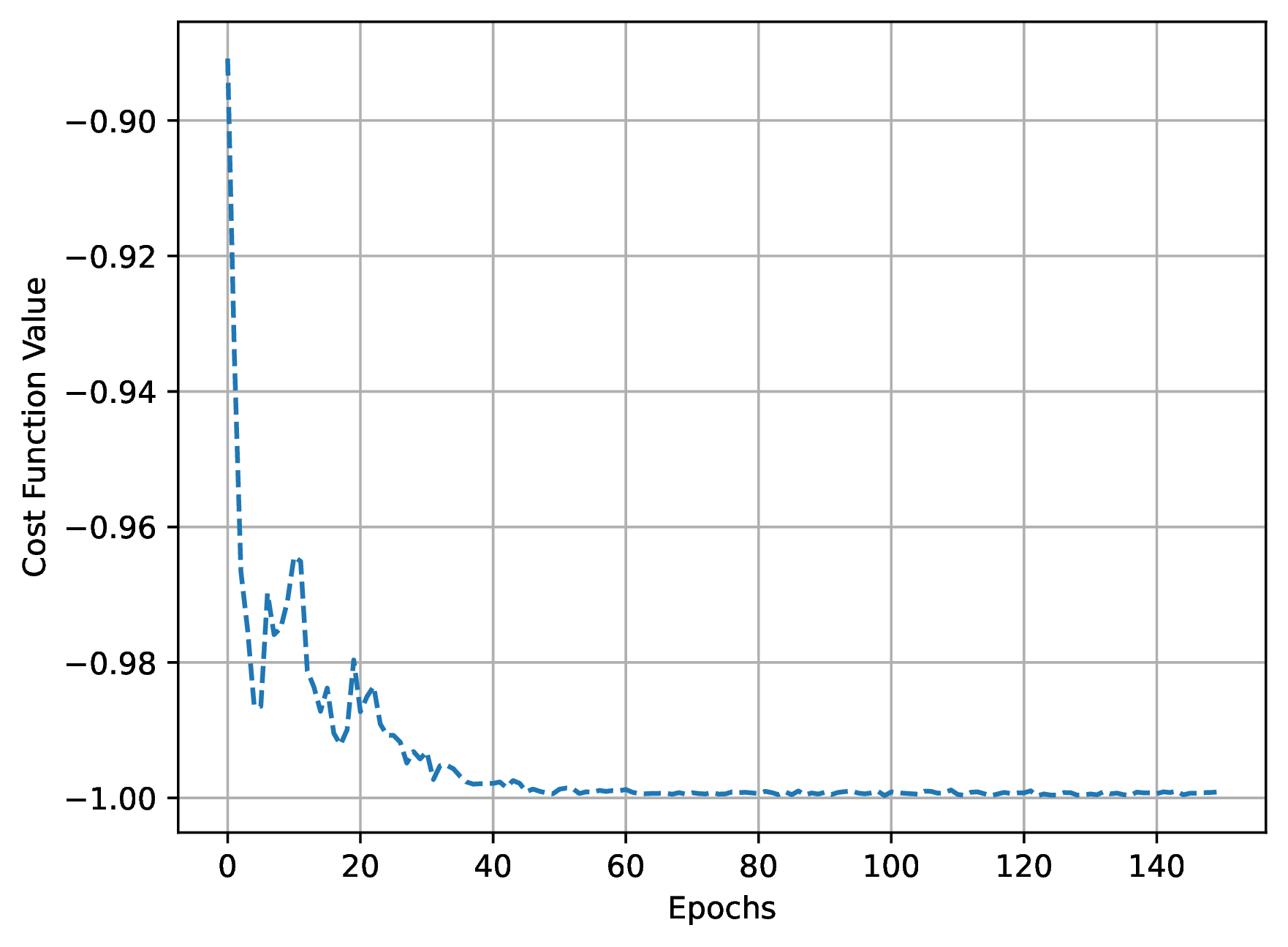}}
    \subfigure[Dictionary = 20, 3-word]{\includegraphics[width=0.3\textwidth]{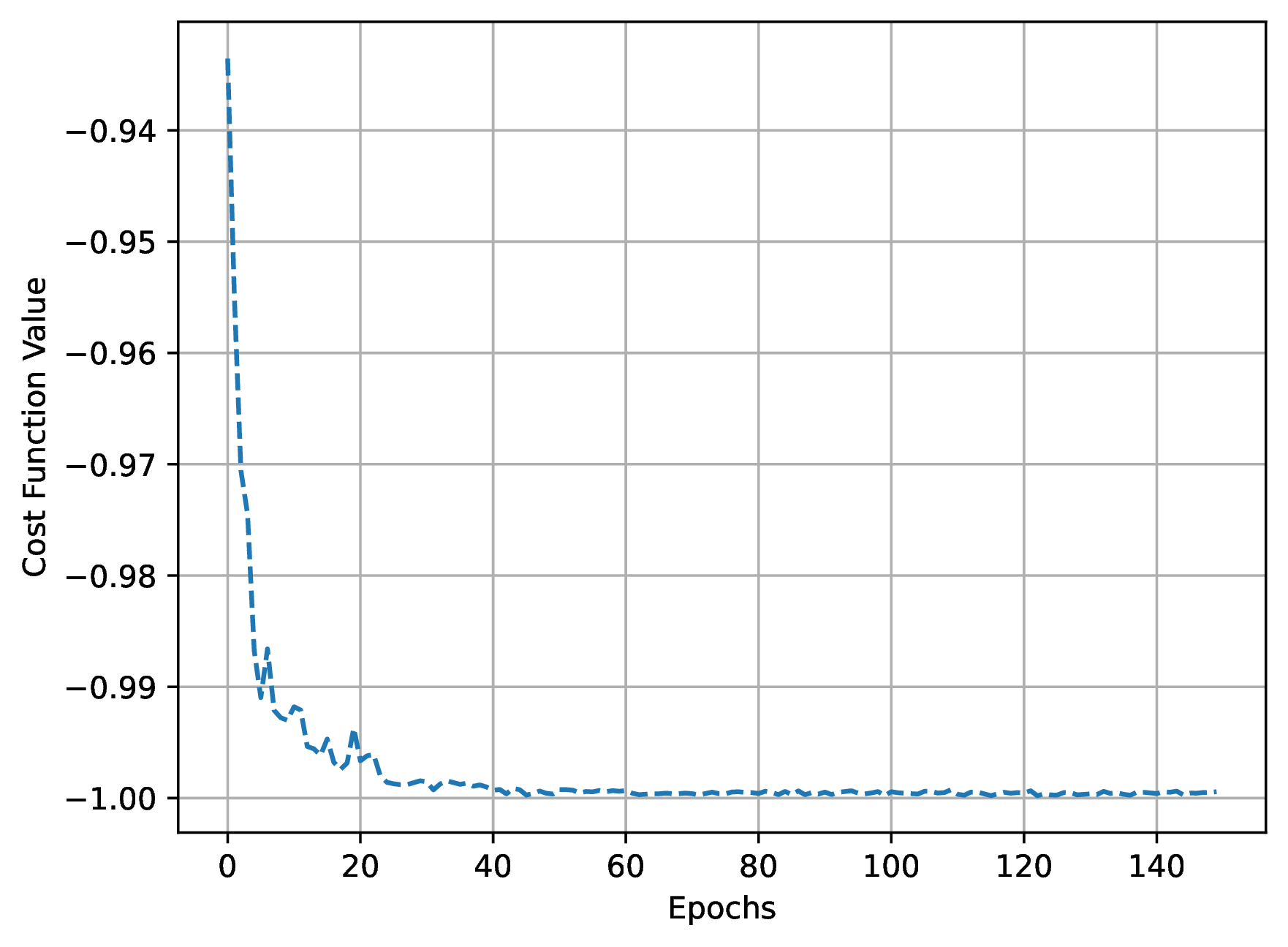}}
    \subfigure[Dictionary = 40, 3-word]{\includegraphics[width=0.3\textwidth]{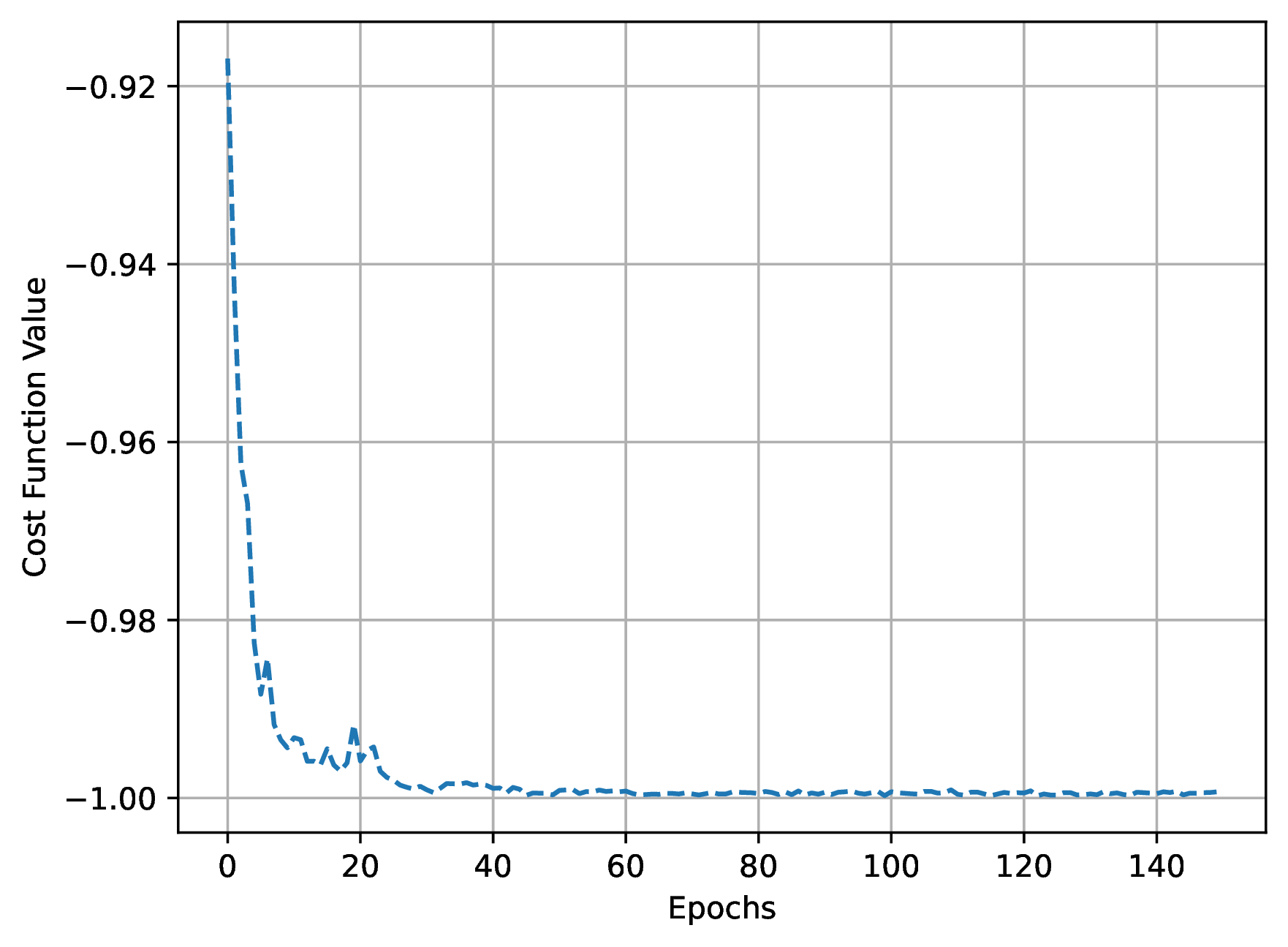}}
    \subfigure[Dictionary = 10, 4-word]{\includegraphics[width=0.3\textwidth]{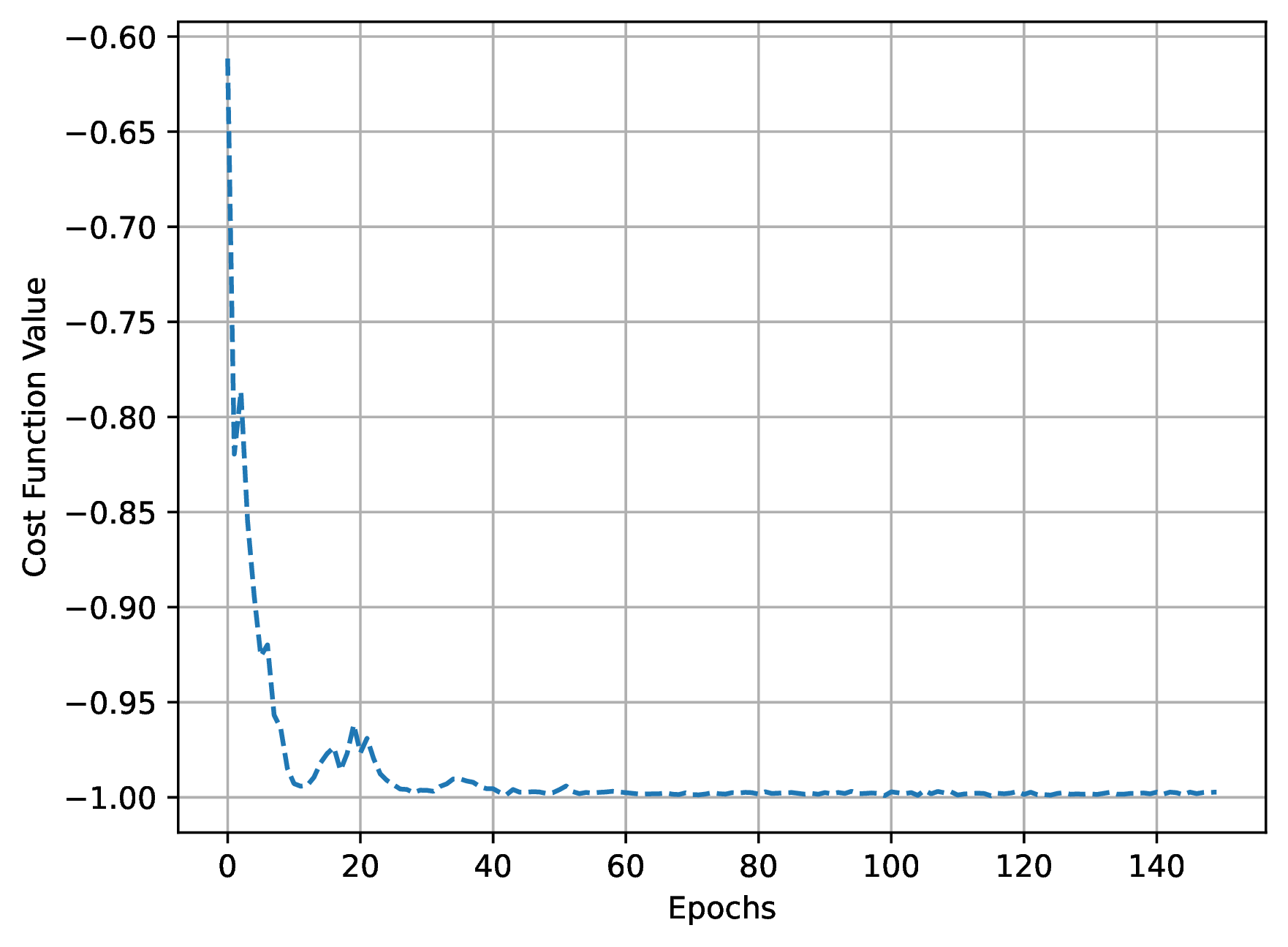}}
    \subfigure[Dictionary = 20, 4-word]{\includegraphics[width=0.3\textwidth]{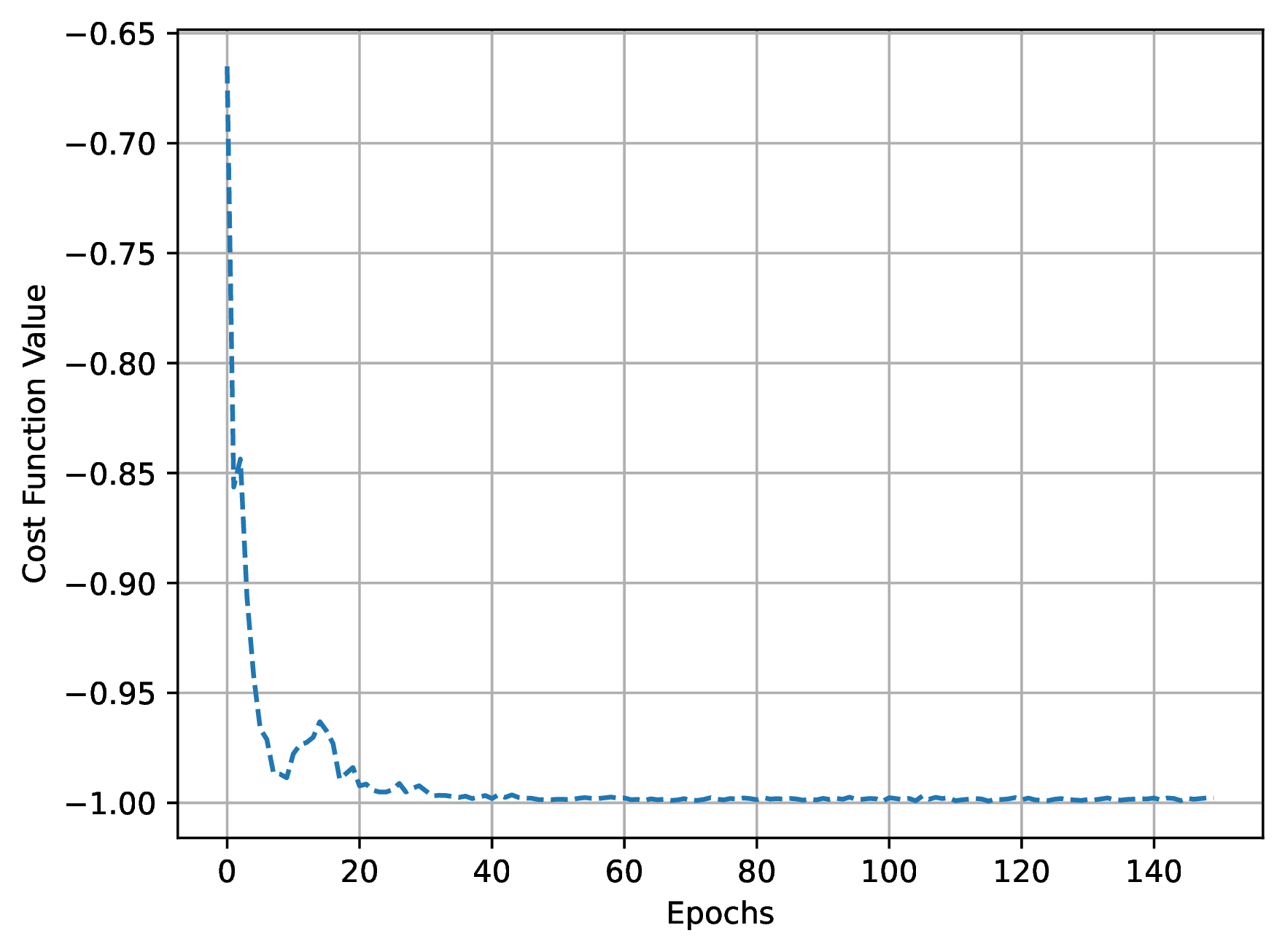}}
    \subfigure[Dictionary = 40, 4-word]{\includegraphics[width=0.3\textwidth]{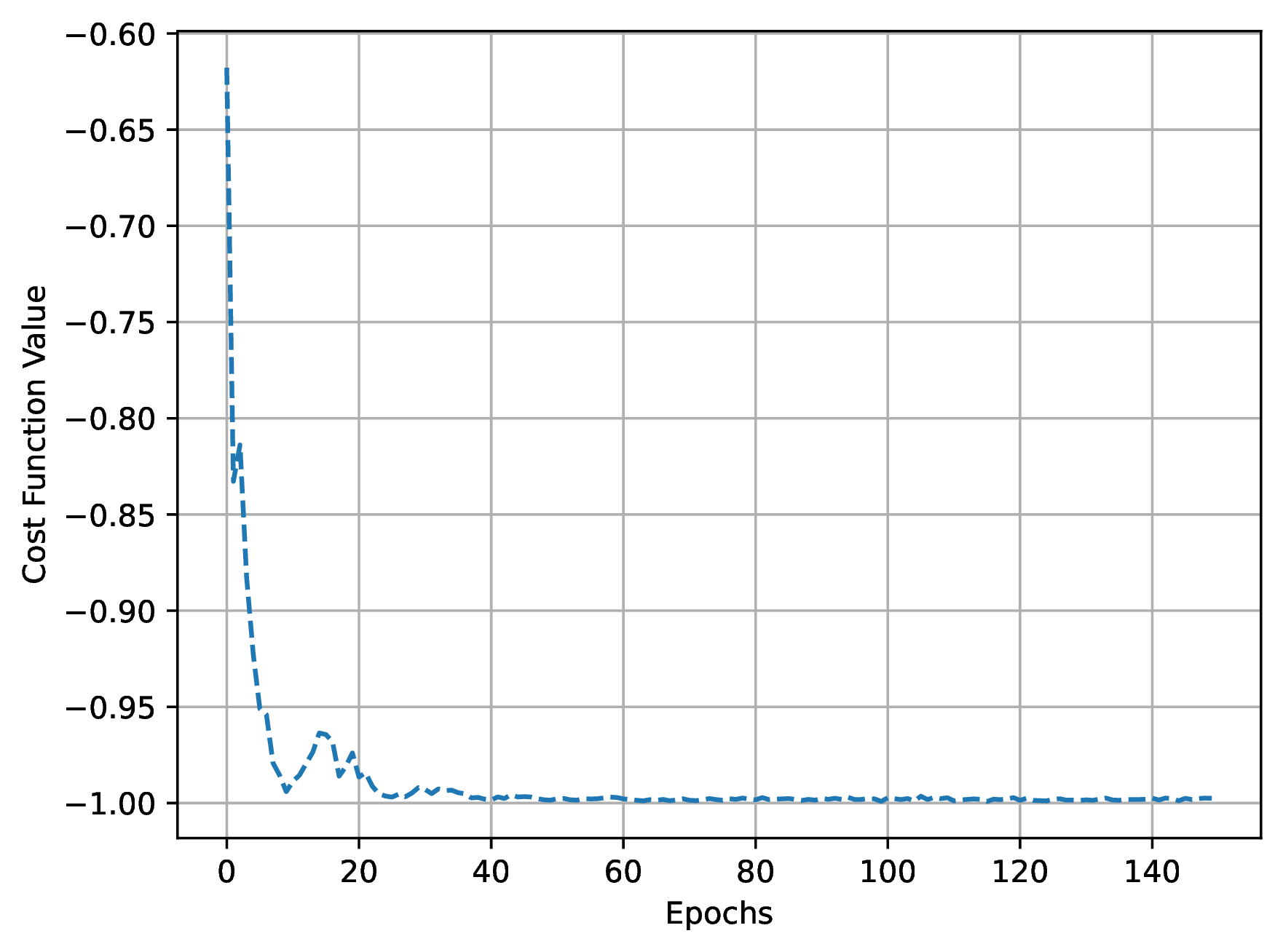}}
    \subfigure[Dictionary = 10, 5-word]{\includegraphics[width=0.3\textwidth]{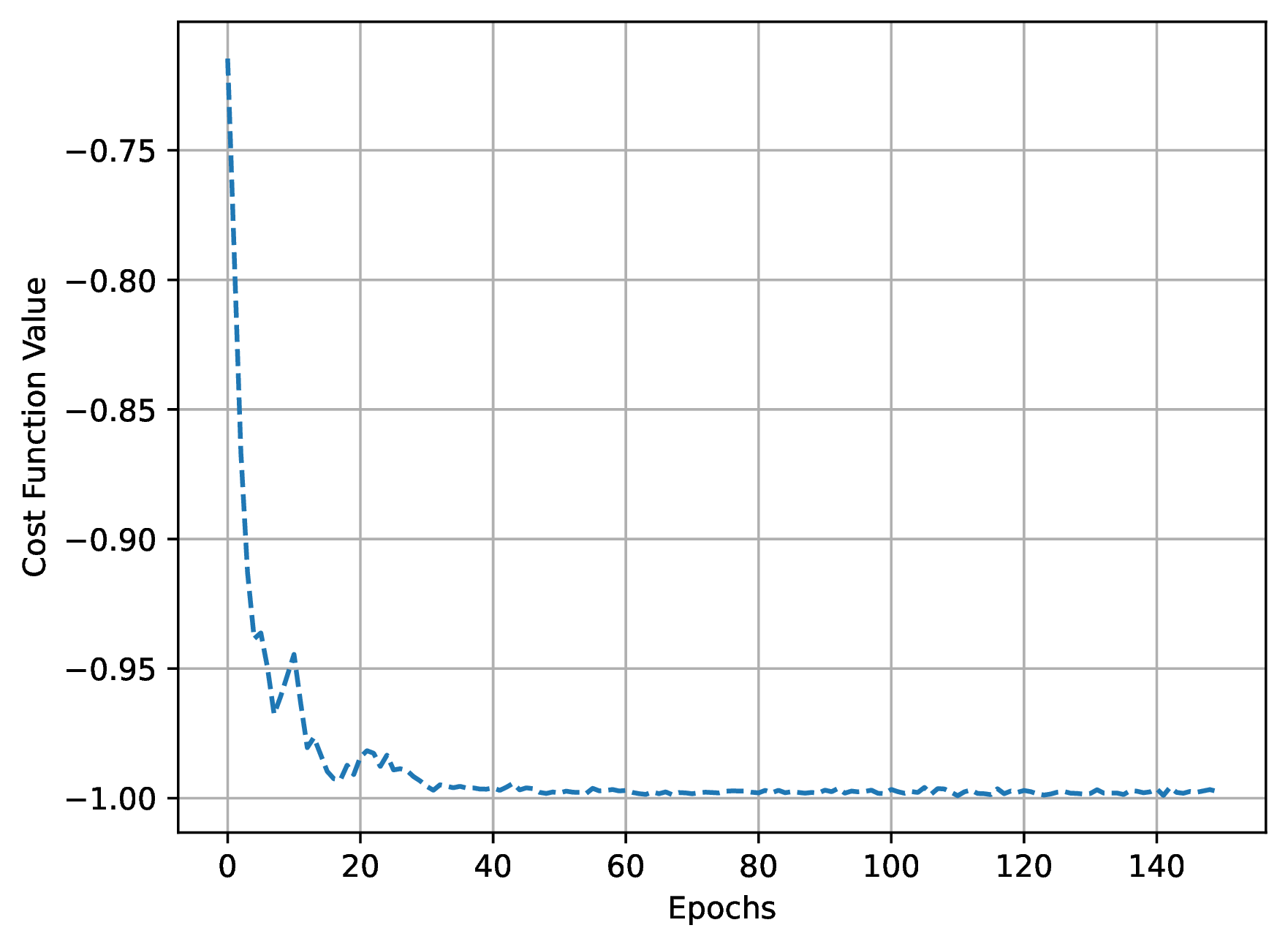}}  
    \subfigure[Dictionary = 20, 5-word]{\includegraphics[width=0.3\textwidth]{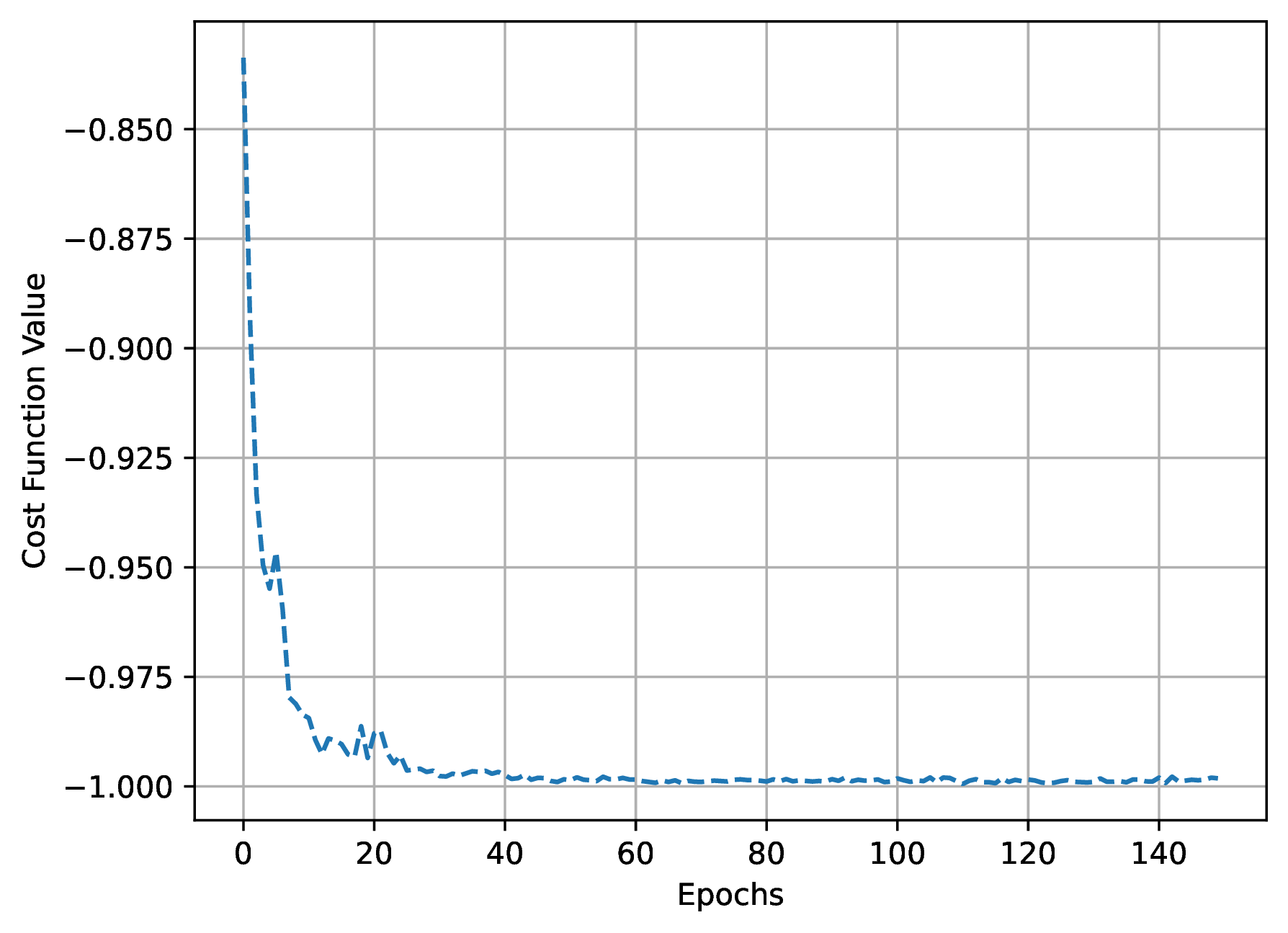}} 
    \subfigure[Dictionary = 40, 5-word]{\includegraphics[width=0.3\textwidth]{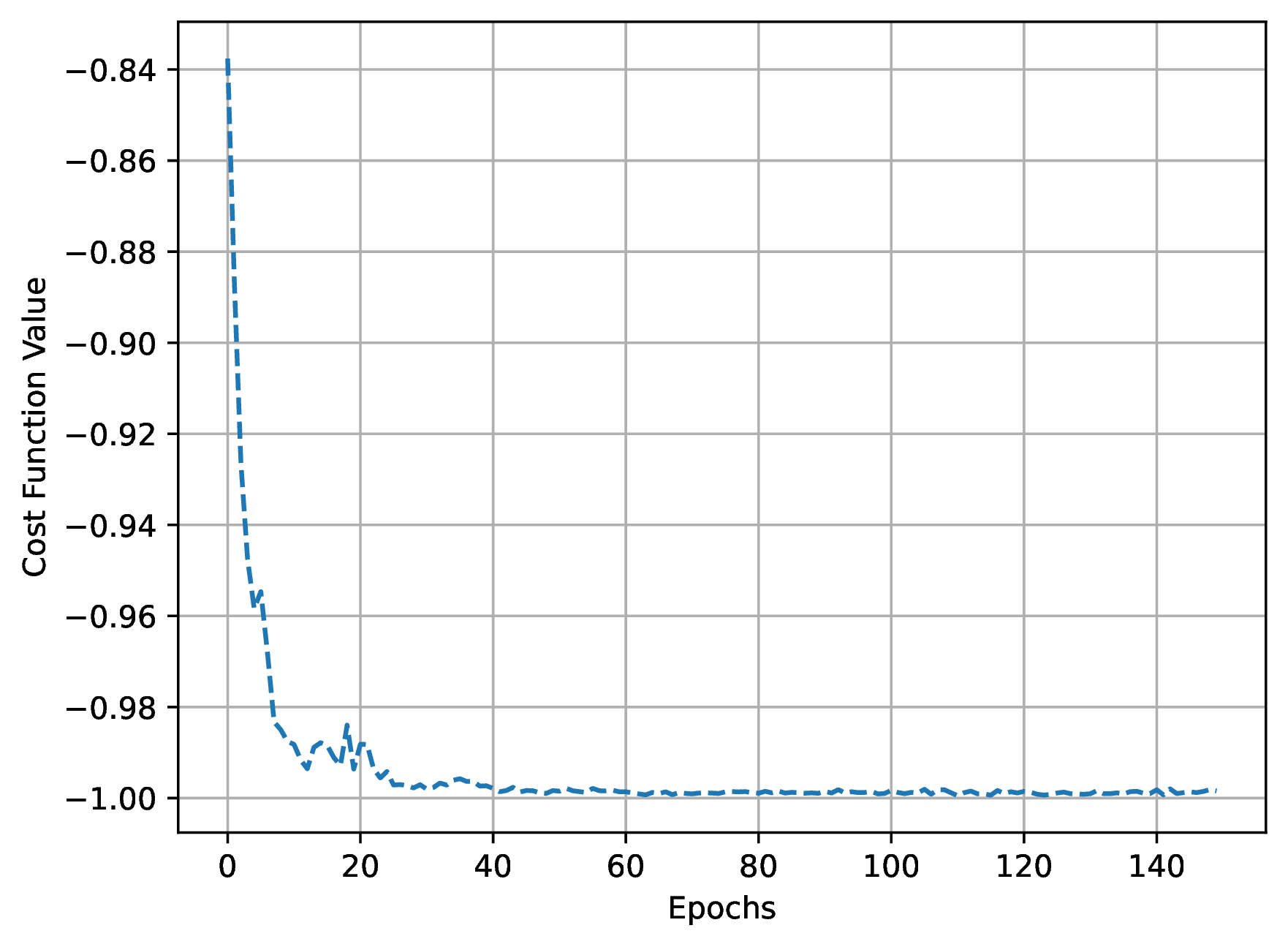}}   
    \caption{Cost function value for all dictionary sizes and sentence lengths}
    \label{fig:cost_function_IH}
\end{figure}

However, we cannot attribute the low accuracies to QGRNN, as the cosine similarity of the numerical values corresponding to the words is nearly perfect for all cases, as shown in Table \ref{tab:accuracy_cosine}. This indicates that QGRNN maintains a strong prediction ability even as dictionary size and sentence length increase. Further supporting this observation, the cost function plot provides additional evidence that the retrieved numerical values are close to their expected values, as shown in Figures \ref{fig:cost_function_IH}. 

As this method currently performs very well for small to moderate dictionary sizes, it is beneficial for specific applications where perfect secrecy is required for usually short messages, such as military and battlefield communication. 


\begin{table}[h!]
\centering
\caption{Accuracy and Cosine Similarity for Different Cases}
\begin{tabular}{llll}
\toprule
\textbf{Dictionary Size} & \textbf{Sentence Length} & \textbf{Accuracy (\%)} & \textbf{Cosine Similarity} \\
\midrule
10 & 3 & 100 & 0.999 \\
10 & 4 & 100  & 0.999 \\
10 & 5 & 100  & 0.999 \\
20 & 3 & 100  & 0.999 \\
20 & 4 & 100  & 0.999 \\
20 & 5 & 80  & 0.999 \\
40 & 3 & 66.67  & 0.999 \\
40 & 4 & 50  & 0.999 \\
40 & 5 & 40  & 0.999 \\
\bottomrule
\end{tabular}
\label{tab:accuracy_cosine}
\end{table}


\section{Conclusion and Future Work}\label{sec:conclusion}

In this work, we explore the potential of QGRNNs for classical data processing and information hiding. We demonstrated that QGRNN effectively reconstructs node features in classical datasets by encoding Iris and MNIST data as graphs and predicting their features. Our results showed that QGRNN’s predictions were very close to actual values, leading to high classification accuracy when tested with classical machine learning models. Building on these results, we introduced a novel information hiding technique that embeds secret messages in graphs, where retrieval is only possible if one has access to time-evolved states and the initial states used to evolve the system. We conducted experiments in which we embedded secret messages of varying lengths (3, 4, and 5 words) and dictionary sizes (10, 20, and 40 words), demonstrating that even with higher complexity graphs, QGRNN achieves high retrieval accuracy with only slight performance decline. These findings suggest that QGRNN can serve as a novel quantum-based information hiding technique that offers enhanced security through quantum constraints such as the no-cloning theorem. The no-cloning theorem ensures that no one can copy the states needed to retrieve the secret message. Overall, this work establishes QGRNN as a powerful tool for processing classical data in quantum systems and introduces its potential for secure information hiding. As quantum technologies continue to advance, further research will be crucial for realizing the full potential of QGRNN in real-world applications.


While this work demonstrates the effectiveness of QGRNN for classical data processing and information hiding, several research directions remain open for further exploration.
\begin{itemize}

\item{Information Hiding:}
In our information hiding technique, we have not explored the use of quantum teleportation to securely transmit the relevant states to the receiver. Future work could investigate how integrating quantum teleportation into our approach enhances security and ensures reliable state reconstruction at the receiver’s end. In addition,  its application in the distribution of symmetric keys for secure communication can be explored further. While our study focused on embedding words in nodes, future work could explore alternative encoding strategies, such as incorporating edges for redundancy. In addition, incorporating error correction mechanisms could further enhance the robustness of the technique.  Furthermore, edges could be utilized to encode additional words within smaller graphs, potentially improving efficiency while maintaining accuracy.

\item{Quantum Sensing:}
Our methodology, combined with classical machine learning, can be leveraged for quantum sensing as a post-processing technique. If quantum states are measured at different times and are known to be classifiable, QGRNN can first reconstruct the relevant features, followed by classification using classical machine learning models. This approach could enable the detection and identification of quantum-sensed data.

\item{Model Extension:}
To better resemble real-world scenarios, future work could evaluate QGRNN under different noise models. Studying its performance in noisy environments would provide insight into the robustness and feasibility of the model for practical quantum computing applications.
\end{itemize}

\bibliographystyle{IEEEtran}
\bibliography{ref}
\end{document}